\documentclass[aps,prd,showpacs,superscriptaddress,nofootinbib,10pt]{revtex4-2}
\usepackage{acronym}
\usepackage{amsfonts}
\usepackage{amsmath}

\usepackage{amssymb}
\usepackage{bm}
\usepackage{cases}
\usepackage{color}
\usepackage{comment}
\usepackage[dvipsnames]{xcolor}
\usepackage{enumerate}
\usepackage{epsfig}
\usepackage{etoolbox}
\usepackage{fancyref}
\usepackage{fancyhdr}
\usepackage[T1]{fontenc} 
\usepackage{graphicx}
\usepackage{hyperref}
\usepackage{indentfirst}
\usepackage{latexsym}
\usepackage{mathrsfs}
\usepackage{mciteplus}
\usepackage{multirow}
\usepackage{rotating}
\usepackage{scrextend}
\usepackage{soul}
\usepackage{subfigure}
\usepackage{verbatim}
\usepackage{amsthm} 
\usepackage{tabularx}
\usepackage{enumitem}    
\usepackage{array}
\usepackage{orcidlink}

\begin{document}

\title{Hamiltonian formalism, master functions and Darboux transformations for perturbed (interiors and exteriors of) nonrotating black holes}

\author{Michele Lenzi\,\orcidlink{0000-0002-0131-2829}
}\email{michele.lenzi@ulb.be}
\affiliation{Physique Théorique et Mathématique, Université Libre de Bruxelles (ULB), Campus Plaine, Building NO, CP 231, Boulevard du Triomphe B-1050, Bruxelles, Belgique}

\author{Guillermo A. Mena Marug\'an\,\orcidlink{0000-0003-3378-9610}
}
\email{mena@iem.cfmac.csic.es}
\affiliation{Instituto de Estructura de la Materia, IEM-CSIC, C/ Serrano 121, 28006 Madrid, Spain}

\author{Andr\'es M\'{\i}nguez-S\'anchez\,\orcidlink{0000-0003-3103-3182}
}
\email{andres.minguez@iem.cfmac.csic.es}
\altaffiliation{Affiliated to the PhD Program, Departamento de F\'isica Te\'orica, Universidad Complutense de Madrid, 28040 Madrid, Spain}
\affiliation{Instituto de Estructura de la Materia, IEM-CSIC, C/ Serrano 121, 28006 Madrid, Spain}

\author{Carlos F. Sopuerta\,\orcidlink{0000-0002-1779-4447}
}
\email{carlos.f.sopuerta@csic.es}
\affiliation{Institut de Ci\`encies de l'Espai (ICE, CSIC), Campus UAB, Carrer de Can Magrans s/n, 08193 Cerdanyola del Vall\`es, Spain}
\affiliation{Institut d'Estudis Espacials de Catalunya (IEEC), Edifici Nexus, Carrer del Gran Capit\`a 2-4, despatx 201, 08034 Barcelona, Spain}

\begin{abstract}
Motivated by their relevance to the interior of nonrotating black holes, classical and quantum Kantowski-Sachs cosmologies have recently attracted increasing attention. This interest has led to the development of a Hamiltonian formalism for axial and polar perturbations, which can be extended to applications in the exterior region. The formalism provides also a description of the background physical degrees of freedom. Moreover, it allows for the construction of all physical perturbative gauge invariants, which can be arranged into canonical pairs associated with master functions. In this work, we review the basis of this Hamiltonian formalism, putting the emphasis on its foundations and fundamental steps rather than on details of the involved calculations. Our discussion focuses on classical and effective aspects, although we also briefly comment on its natural role in the quantization of perturbed black holes. Adopting this formalism we present a geometric interpretation of Darboux transformations between pairs of master functions, characterizing them as generalized canonical transformations that preserve the Hamiltonian structure of the perturbations as harmonic oscillators subject to certain potentials. This bijective correspondence between such canonical transformations and Darboux transformations, which was recently proved to hold for axial perturbations, is here extended to the case of polar perturbations. In addition, we demonstrate the existence of canonical transformations that, similarly to Darboux transformations, mix axial and polar master functions. 


\end{abstract}

\maketitle

\section{Introduction \label{Intro}}

Black hole (BH) spacetimes are some of the most fascinating entities in gravitation and astrophysics \cite{BHs,BHM}. They constitute a fundamental challenge in classical gravitational theories such as general relativity, where in spite of their simplicity and feasibility, they lead to the occurrence of strong singularities in their interior where the predictability of physics breaks down \cite{BHCl}. In semiclassical gravity, it is necessary to build models that avoid these singularities by the presence of a quantum matter content while keeping the classical nature of the spacetime, all of this without recurring to unjustified fine tunings \cite{BHSemiCl}. Moreover, if gravity also admits a quantum description, one must explain how to circumvent other fundamental problems like unitarity or the information paradox \cite{BHQu}.

To understand the behavior of BHs and the physical phenomena they can generate, significant attention has been devoted over the years to the study of their perturbations \cite{RW,Teu,BHM}, considering both their geometry and the possible existence of matter fields around them. This perturbative analysis is crucial to elucidate whether BH solutions are stable and therefore physically significant. This stability is particularly important to understand the fate of classical singularities or their radiation mechanisms \cite{Nollert,Hawking}. The latter is relevant if one wants to consider the gravitational emission of BHs that are in the process of reaching their equilibrium state. This is a situation with practical applications in astrophysics, for instance after the merging of two compact objects that creates a BH. Typically, after the merger and a subsequent transition stage, the BH emits gravitational waves as its excitations decay during the ringdown phase \cite{QNMs}. This gravitational radiation is composed of some characteristic constitutents that are called quasinormal modes \cite{Nollert,QNMs}, which carry information about the BHs features and, in principle, also of the gravitational interaction theory. This allows for tests of general relativity using data from the gravitational waves produced by mergers during ringdown, and opens fascinating fronts for the study of the gravitational field in strong regimes \cite{Tests}.

A complication in the investigation of perturbations in general relativity, and therefore for BH solutions, comes from the fact that one must compare the unperturbed spacetime with perturbed versions of it. In particular, the points of the original manifold have to be identified with those of the perturbed manifold, and this introduces a freedom that complicates the analysis. The problem was already present in the discussion of gravitational waves around flat spacetime that Einstein carried out \cite{EinsteinWave} soon after publishing his theory of general relativity, and that he realized to be associated to the freedom in performing infinitesimal, perturbative diffeomorphisms. In the same way that Einstein was able to isolate the part of the gravitational wave solutions that did not depend on that freedom, the physically relevant information about the perturbations of BHs is contained only in quantities that are invariant under changes in such an identification of points. These quantities are usually called perturbative gauge invariants \cite{GI,GI2}. At leading perturbative order, they are linear combinations of perturbations that are not affected by perturbative diffeomorphisms. They can be multiplied by any background-dependent factor and remain gauge invariants, since a perturbative diffeomorphism would produce a perturbation of the background that would subsequently lead to quantities that are not linear in the perturbations any longer, and hence ignorable at the considered leading order. Perturbative gauge invariants encode the physical degrees of freedom of the perturbations, with all other information about them corresponding to pure gauge.

In this work, we are going to focus our discussion on nonrotating BHs, with orbits of spherical symmetry. Although many of the analysis that we are going to present apply as well to the more general situation of rotating BHs, the use of spherical symmetry will allow us to considerably simplify our calculations. Each of these symmetry orbits forms a two-dimensional manifold with the topology of a two-sphere, $S^2$. The set of these orbits, on the other hand, forms as well a two-dimensional manifold, $\Sigma_2$, in this case of Lorentzian signature. The spacetime symmetry allows to describe many of our fields in a two-dimensional reduced version, as functions on $\Sigma_2$. In particular, perturbative gauge invariants are often described by functions of this type, namely scalar fields on $\Sigma_2$. Master functions \cite{CPM1,Zer,Mon} are scalar fields of this kind that satisfy second-order differential equations on this two-dimensional Lorentzian spacetime, with a differential part corresponding to the Laplace-Beltrami operator associated with its induced metric. During ringdown, the spacetime region between the BH horizon and the spatial region at infinity contains perturbations that must be {\sl expelled} to reach equilibrium. The solutions to the wave equation of the master functions with outgoing boundary conditions at infinity and also at the horizon, with the normal pointing towards its interior, are in the classical treatment known as quasinormal modes \cite{Nollert}.   
 
There exists an intriguing relation between different perturbative master functions, given by the so-called Darboux covariance \cite{Darboux,CM1}. The origin of this relation was recently investigated in the Lagrangian formalism in Refs. \cite{CM1,Lenzi:2021wpc,CM2}. In general, suppose that $\varphi$ is a master function that satisfies a wave equation in two dimensions for a certain potential $V_{\ell}$ which depends on the angular-momentum number $\ell$ of the perturbations under analysis. We can then consider the Darboux transformation $\Psi=\varphi^{\prime}+g_{\ell} \varphi$, where the prime stands for the derivative with respect to a tortoise coordinate in $\Sigma_2$, and such that $g_{\ell}$ satisfies the Riccati equation $g_{\ell}^{\prime}-g_{\ell}^2+V_{\ell}+c_{\ell}=0$, with $c_{\ell}$ being a constant. Then $\Psi$ is a solution of a wave equation with a new potential $v_{\ell}=V_{\ell}+2 g_{\ell}^{\prime}$. We will describe these transformations in more detail later in our discussion, but from the brief comments above we can already understand that these transformations between master functions can be used to relate the physics associated with each of them. In fact, they have been used in Schwarzschild BHs to relate the quasinormal modes corresponding to different master functions, showing that their frequencies must coincide \cite{BHM,Chandra,Yurov, Glampe}. This property is known as isospectrality.

Most of the studies of perturbations of BHs have been carried out using a Lagrangian formalism, or equivalently starting with the Einstein-Hilbert action (with appropriate boundary terms) of general relativity in Lagrangian form \cite{Wald}. Nonetheless, the problem can be equally analyzed in a Hamiltonian framework, and in fact the Hamiltonian formulation can be advantageous for certain issues, like the counting of physical degrees of freedom, the identification of gauge symmetries or, partly related to the latter, the direct identification of perturbative gauge invariants. Actually, in a Hamiltonian formalism, the generators of perturbative diffeomorphisms are constraints that appear from the linearization of the diffeomorphism constraints of general relativity. At leading order, perturbative gauge invariants are simply linear combinations of perturbations that commute with those perturbative constraints, so that the transformations they generate leave invariant the considered combinations. In addition, a Hamiltonian description makes the passage to the quantum theory much more manageable. 

A Hamiltonian interpretation of Darboux transformations was discussed in Ref. \cite{MatveevSalle} in a different context, namely in connection with the Korteweg–De Vries equation. However, our interest in this work is focused on gravitational perturbations. The Hamiltonian description of the leading-order perturbations of nonrotating BHs was already investigated in the 2000's in Refs. \cite{JMMG1,DC}, together with a Lagrangian formulation for higher-order perturbations \cite{Brizu,Brizu2}. Those works used extensively the splitting of the four-dimensional spacetime between spherical orbits $S^2$ and the two-dimensional set of those symmetry orbits, $\Sigma_2$. In particular, the spherical symmetry was used to expand all four-dimensional fields in harmonic modes on $S^2$, characterized by an angular momentum label $\ell$ and an azimuthal label $m$. In spite of the progress achieved in those works, the simultaneous dependence on two different directions, one in which the background solutions vary, and another in which the dynamical evolution is described, complicated the analysis to a very intricate level. 

Lately, significant attention has been paid to the consideration of the interior of such nonrotating BH spacetimes. This interior can be described as a Kantowski-Sachs (KS) geometry \cite{KS,KS2,KS3}, and therefore treated as a cosmological solution \cite{AOS,AOS2,AO,BAG,BGA}. A Hamiltonian formalism in this interior region eludes the problem of interdependence mentioned above because in this case the direction of evolution is the same in which the background spacetime solutions change. Moreover, corrections to general relativity, for instance owing to quantum geometry phenomena, can be included by effective descriptions while allowing the same general treatment of the perturbations \cite{MM,MM2}. Remarkably, this kind of corrections can remove the essential singularity in the BH interior, solving the possible obstructions to construct a Hamiltonian formalism around it \cite{MM}. Moreover, the Hamiltonian formulation built in this way can be naturally extended from the interior region to the exterior region, either by means of a complex canonical transformation of the background \cite{AOS2} or in terms of some convenient real background functions, permitting an extension of their range \cite{MGAC}.

In recent studies, indeed, a complete Hamiltonian description of nonrotating perturbed BHs has been successfully attained, including in it both the background and its geometric perturbations, as well as the interior and exterior regions \cite{BGA,MM,MM2,MGAC,MGA}. One of the aims of this work is to concisely review the results of those studies, avoiding unnecessary formulas that may obscure the physically relevant points of the discussion and explaining the basis of the formalism. In this review, we will focus our attention on classical and effective aspects, with only some comments about the road to quantization of the perturbed BHs. In particular, we want to show in a clear and simple way how Darboux transformations appear in our Hamiltonian formalism as a subclass of canonical transformations with certain properties. In this respect our work goes beyond the revision of previous analyses and contains some new results about Darboux transformations. In more detail, for perturbations of polar (or even) parity (i.e., which transform under parity transformations like the scalar spherical harmonics), we discuss for the first time the one-to-one correspondence between canonical transformations preserving the generalized harmonic-oscillator form of the quadratic perturbative Hamiltonian and the Darboux transformations, a correspondence that had already been demonstrated for axial (or odd) perturbations in Ref. \cite{MGAC}. We also show the canonical counterpart of the Chandrasekhar transformation that mixes polar and axial polarities \cite{BHM}, and comment about some simple generalizations of it. 

The rest of this work is organized as follows. In Sec. \ref{SecII} we review the Hamiltonian formulation of the background BH spacetime, adapted to the interior region, but with a natural extension to the exterior that we discuss. We then summarize the Hamiltonian description of the perturbed BH, including axial and polar perturbations, in Sec. \ref{SecIII}. There, we also describe the basis of modes in which we expand all the perturbations of the geometry. We then explain in Sec. \ref{SecIV} how to perform a canonical transformation in our gravitational system that isolates the perturbative gauge invariants, describing them by canonical pairs of perturbative mode variables. The final result of this transformation is a canonical set of variables for the background and the perturbations in which the perturbative gauge constraints appear as a subset, and that is subject to only a nontrivial constraint, namely a total Hamiltonian, which contains both background and quadratic perturbative contributions. In Sec. \ref{SecV} we introduce an additional canonical transformation that allows us to express the perturbative gauge invariants in terms of the familiar master functions employed in BH perturbation theory. The relation between Darboux transformations and canonical transformations is discussed in Sec. \ref{SecVI}. We show that Darboux transformations that preserve polarity are in one-to-one correspondence with canonical transformations respecting the diagonal form of the perturbative Hamiltonian, avoiding unphysical nonlocality, and with momenta that are equal to the Hamiltonian derivative of the configuration variables. This result had been demonstrated in Ref. \cite{MGAC} for the axial sector, and it is here extended to polar perturbations. Moreover, we also investigate canonical transformations corresponding to Darboux transformations that violate parity. Our conclusions and some lines of further research are presented in Sec. \ref{SecVII}. Finally, two appendices are added with some details of the calculations and explicit formulas.

In the rest of our discussion, we choose units such that the speed of light, Newton's constant, and Planck's reduced constant are set equal to one. We use lowercase Latin indices from the middle of the alphabet for indices on the sections with constant value of the Hamiltonian evolution parameter, and capital Latin indices from the beginning of the alphabet for indices on the two-sphere $S^2$. 

\section{Background \label{SecII}}

The metric in the interior of a nonrotating, uncharged BH adopts a KS form, and can be written in terms of triad variables as follows \cite{AOS,AOS2,AO,BGA}:
\begin{equation}
    \label{metric}
    \text{d}s^2 = \frac{p_b^2}{L_0^2}\bigg(-|p_c| \tilde{N}^2 \text{d}\chi^2 + \frac{1}{|p_c|} \text{d}\zeta^2\bigg) + |p_c|(\text{d}\theta^2 + \sin^2\theta\text{d}\phi^2).
\end{equation}
Here, $\theta$ and $\phi$ are angular coordinates on the unit two-sphere, and $\chi$ and $\zeta$ are coordinates on the set of spherical orbits $\Sigma_2$. We have chosen this notation to avoid the unconscious identification of one of them with a time and the other with a spatial radial coordinate. In the KS region where $p_b$ is real, and hence its square is positive, $\chi$ is a timelike coordinate. More generally, we will treat it as the coordinate in the direction of Hamiltonian evolution, independently of whether this direction is timelike or spacelike, in the event that $p_b^2$ becomes negative, as we discuss below.

The metric functions $p_b$ and $p_c$ are assumed to depend exclusively on $\chi$, and have canonical momenta given by extrinsic-curvature variables $b$ and $c$, respectively, chosen such that \cite{AOS,AOS2,AO}
\begin{equation}
\label{backgroundbrac}
\{b,p_b\}=1,\qquad \{c,p_c\}=2.
\end{equation}
To avoid infrared issues related to infinite volumes in the sections with constant value of the coordinate $\chi$, we consider that the $\zeta$-direction has been compactified, with $L_0$ serving as a compactification parameter that can also be used to pass to a suitable non-compact limit at the end of all the calculations. 

We note that $p_b$ and $p_c$ have a clear geometric meaning. Since our background is spherically symmetric, we can use this symmetry to identify $|p_c|$ as the square conformal factor of the corresponding two-dimensional orbits (if the $\zeta$-direction were timelike, this is what we would call the radial function). On the other hand, $p_b$ is the scale factor of the two-dimensional submanifold $\Sigma_2$ that provides the set of all those symmetry orbits. The splitting of our four-dimensional manifold into two two-dimensional submanifolds (spherical orbits and the set of them) will prove useful in our analysis. In this splitting, as we have already pointed out, the respective spacelike or timelike character of the coordinates $(\chi,\zeta)$ on $\Sigma_2$ depends on whether $p_b$ is real (so that its square is positive) or imaginary (and hence its square is negative). In this respect, we note that the complex canonical transformation \cite{AOS2,MGAC,MGA}
\begin{equation}
    \label{complexsymplec}
    \bar{b} = -ib, \qquad \bar{p}_b = ip_b, \qquad \bar{c} = c, \qquad \bar{p}_c = p_c,
\end{equation}
interchanges the spacelike and timelike roles of $\chi$ and $\zeta$. Thus, this transformation can be understood as relating the interior geometry of the BH with its exterior correspondent. For functions on phase space that depend only on the canonical pair $c$ and $p_c$ and on functional combinations of $b$ and $p_b$ that remain real under the considered transformation, for example $p_b^2$ and $bp_b$, the only effect is modifying their real range. Since our analysis involves only functions of this type (as is the case for the metric), we can adopt a unified description and always use the notation without bars, taking into account the allowed range of variation when specializing to the interior or exterior.

The KS background geometry is subject only to a Hamiltonian constraint, which is given by \cite{AOS,AOS2,BGA}
\begin{equation}
\label{backgroundHam}
\tilde{H}_{\text{KS}}[\tilde{N}] = -\tilde{N}\frac{L_0}{2}\bigg( \Omega_b^2 + \frac{p_b^2}{L_0^2} + 2\Omega_b\Omega_c \bigg).
\end{equation}
We have introduced the background functions $\Omega_a = a p_a/L_0$ ($a=b,c$), which are simply the generators of dilations in the $b$-- and $c$--sectors of the phase space (for $a=b$ and $c$ respectively). Remarkably, the above Hamiltonian depends only on functions of the $b$--sector that in fact remain real under the complex canonical transformation \eqref{complexsymplec}. 

For the specific case of the Schwarzschild interior solution in classical general relativity, the background variables and the lapse in our canonical formulation adopt the following expressions \cite{MGAC}:
\begin{equation}
\label{Schwarz}
\frac{p_b^2}{L_0^2} = - \tilde{N}^{-1}= -\chi^2 f, \qquad |p_c| = \chi^2, \qquad \Omega_b = \chi f, \qquad \Omega_c = M,
\end{equation}
where $f(\chi) = 1 - 2M/\chi$ and $M$ is the (Arnowitt-Deser-Misner or ADM \cite{ADM}) mass. These expressions are on-shell, that is, they only hold on solutions. Furthermore, our choice of lapse reproduces the familiar Schwarzschild coordinate choice for the {\sl interior time}, which in the exterior region becomes the radial coordinate. These two regions correspond to the intervals $\chi< 2M$ and $\chi>2M$, respectively, where the function $f$, and hence $p_b^2$, is positive or negative definite. The variable $\Omega_b$ remains real in both cases, in full agreement with our previous comments.  

\section{Perturbations \label{SecIII}}

As we have already discussed, we consider compact sections of constant evolution parameter $\chi$ which have the topology of $S^1 \times S^2$. Thanks to this compact topology, zero modes are isolated and can be treated exactly. On the other hand, employing the symmetry of our background, we can expand our perturbations in real spherical harmonics and Fourier modes in the $\zeta$-direction. We recall that this is a Killing direction of the background spacetime.

For the spherical part, we use a (real) Regge-Wheeler-Zerilli basis of harmonics (see e.g. Refs. \cite{DC,Brizu,MM}). Spherical harmonics are labeled by their angular momentum $\ell$, which takes nonnegative integer values, and the azimuthal number $m$, which is also an integer and ranges from $-\ell$ to $\ell$. These harmonics split in polar and axial contributions under parity transformations. A polar harmonic of eigenvalue $-\ell(\ell+1)$ for the Laplace-Beltrami operator on the unit two-sphere $S^2$ has parity eigenvalue equal to $(-1)^{\ell}$. Scalar harmonics $Y_{\ell}^{m}$ are all polar.

We then decompose any symmetric tensor on a $\chi$-constant section as
\begin{equation}
T_{ij}\text{d}x^i \text{d}x^j=T_{\zeta\zeta} \text{d}\zeta^2+2 T_{\zeta A}\text{d}\zeta \text{d}x^A+T_{AB} \text{d}x^A \text{d}x^B .
\end{equation}
For scalars on $S^2$ (for example, the component $T_{\zeta\zeta}$ above), we adopt an expansion on scalar harmonics $Y_{\ell}^{m}$. The coefficients of this expansion inherit the labels $\ell$ and $m$ of the harmonics, and are further Fourier expanded with respect to their dependence on $\zeta$. Each Fourier mode, as usual, is associated with an integer $n$, with a frequency equal to $\omega_n=2\pi |n|/L_0$. The resulting coefficients will therefore be characterized by the three integers $\mathfrak{n}=(n,\ell,m)$. In our analysis, we are going to absorb the zero-mode contribution directly on the background variables, since we want to treat them exactly at the order of our perturbation truncation, as we commented above. In other words, our mode expansions will not include zero modes. On the other hand, note that, at the end of the day, the coefficients of our mode expansions only depend on the coordinate $\chi$, and in this sense are dynamical, inasmuch as the evolution on this coordinate is described by Hamiltonian methods. These dynamical coefficients are going to play the role of variables for the perturbations of the background metric. 

For a covector on $S^2$, the harmonic expansion is made in terms of a basis of vector harmonics, $\{X^m_{\ell\;A},Z_{\ell\;A}^m\}$. More precisely, the harmonics $Z^m_{\ell\;A}$ are obtained from the scalar spherical harmonics by taking their covariant derivative with respect to the metric $\gamma_{AB}$ of the unit two-sphere, and therefore have the same polar behavior as the scalar harmonics. By contrast, the harmonics $X_{\ell\;A}^m$ are reached by contracting this covariant derivative with the antisymmetric tensor on $S^2$, which changes the polarity from polar to axial. Since the covariant derivative of the scalar harmonic for $\ell=0$ vanishes, vector harmonics contribute only for $\ell\geq 1$.

Following the same reasoning, tensor harmonics are obtained by taking one additional covariant derivative on $S^2$. Starting from $X_{\ell\;A}^m$ and symmetrizing the resulting indices on $S^2$, we get the axial symmetric tensor harmonics $X_{\ell\;AB}^m$. In the polar case, we can instead symmetrize two covariant derivatives acting on the scalar spherical harmonics. In this case, it is convenient to decompose the result into its trace and traceless parts. The trace is defined by contracting the spherical indices with the metric on $S^2$. In this way, we obtain the polar traceless symmetric tensor harmonics $Z_{\ell\;AB}^m$, while the trace part of the polar symmetric tensor harmonics is expanded using $\gamma_{AB}$, thereby completing the tensor harmonic basis with $Y_{\ell\;AB}^m\propto \gamma_{AB} Y_{\ell}^m $.  
  
To avoid having to deal with complex relations between the variables that describe the perturbations in our mode expansions, we choose to work with a real basis of spherical harmonics and of Fourier modes. For the real scalar harmonics, from which all other harmonics are constructed, we take \cite{MM}
\begin{equation}
Y_{\ell}^m=\,_{c}Y_{\ell}^m \quad {\rm if} \; m=0, \quad (-1)^m \left(\,_{c}Y_{\ell}^m + \,_{c}Y_{\ell}^{m \,\star}\right)/\sqrt{2}
\quad {\rm if} \; m>0, \quad  (-1)^m \left(\,_{c}Y_{\ell}^{|m|} - \,_{c}Y_{\ell}^{|m|\,\star} \right)/(i \sqrt{2}) \quad {\rm if} \;m<0,
\end{equation}
where $\,_{c}Y_{\ell}^m$ stands for the standard complex spherical harmonics, and the star symbol denotes complex conjugation. Similarly, for the Fourier expansion, we employ the real modes
\begin{equation}
F_0=1/\sqrt{L_0},\quad F_{n}=\sqrt{2/L_0} \cos{(\omega_n \zeta)} \quad {\rm if} \; n>0, \quad F_{n}=\sqrt{2/L_0} \sin{(\omega_n  \zeta)} \quad {\rm if} \; n<0.
\end{equation}
Alternative, one can instead use complex spherical harmonics and Fourier plane waves, but then one must take into account that the mode coefficients are also complex, though related in pairs by suitable complex conjugation relations in order that the subsequent metric perturbations are real. Certainly, the final result is independent of the procedure chosen.

Having explained how perturbations are decomposed, we now denote by $h=h_{ij}\text{d}x^i\text{d}x^j$ the perturbation of the three-metric induced on the $\chi$-constant sections, and by $p=p_{ij}\text{d}x^i\text{d}x^j$ its conjugate momentum, where indices $i,j,\ldots$ are lowered and raised with the unperturbed three-metric. We further denote by $C$ and $B=B_i\text{d}x^i$ the perturbations of the lapse function and the shift vector, respectively, in the 3+1 decomposition of the metric associated with the foliation along the $\chi$-direction. We then introduce our mode expansion that, with a convenient background-dependent scaling of the different perturbations (that takes into account the orthonormal relations satisfied by our mode basis \cite{MM,MM2}), can be expressed as follows:
\begin{equation}
\label{modedecom}
h_{ij} = \sum_{\mathfrak{n}}h^{\mathfrak{n}}_{ij}, \qquad p_{ij} = \sqrt{\text{det}(\gamma)}\sum_{\mathfrak{n}}p^{\mathfrak{n}}_{ij}, \qquad C = \kappa\sqrt{\frac{p_b^2|p_c|}{L_0^2}} \sum_{\mathfrak{n}}C^{\mathfrak{n}}, \qquad B_i = \kappa\sum_{\mathfrak{n}}B^{\mathfrak{n}}_i,
\end{equation}
where $\kappa=16\pi$ in our choice of units, and we have made use of the determinant of the metric of the two-sphere. The sum runs over all possible combinations of labels in $\mathfrak{n}$, excluding the zero modes, as discussed above, as well as those with $\ell\leq1$, which correspond to modes that are physically irrelevant for the present discussion \cite{MM,MM2}. In more detail, employing our mode basis and our compact notation for one-- and two--forms, we have
\begin{equation}
\label{modespertub}
\begin{aligned}
&\begin{aligned}
h^{\mathfrak{n}} &= h_6^{\mathfrak{n}}Y^{m}_{\ell}F_{n} \text{d}\zeta^2 - 2(h_1^{\mathfrak{n}}X^{m}_{\ell\;A}-h_5^{\mathfrak{n}}Z^{m}_{\ell\;A})F_{n} \text{d}\zeta \text{d}x^A + \left(h_2^{\mathfrak{n}}X^{m}_{\ell\;AB} + h_3^{\mathfrak{n}}Y^{m}_{\ell\;AB} + h_4^{\mathfrak{n}}Z^{m}_{\ell\;AB}\right)F_{n} \text{d}x^A \text{d}x^B,
\end{aligned}\\
&\begin{aligned}
p^{\mathfrak{n}} &= \frac{p_b^4}{L_0^4p_c^2}p_6^{\mathfrak{n}}Y^{m}_{\ell}F_{n}\text{d}\zeta^2 - \frac{p_b^2}{\ell(\ell+1)L_0^2}\left(p_1^{\mathfrak{n}}X^{m}_{\ell\;A}-p_5^{\mathfrak{n}}Z^{m}_{\ell\;A}\right)F_{n}\text{d}\zeta \text{d}x^A \\
&+ 2 p_c^2 \left[\frac{(\ell-2)!}{(\ell+2)!}\left(p_2^{\mathfrak{n}}X^{m}_{\ell\;AB} + p_4^{\mathfrak{n}}Z^{m}_{\ell\;AB}\right)+ \frac{1}{4} p_3^{\mathfrak{n}}Y^{m}_{\ell\;AB} \right] F_{n} \text{d}x^A \text{d}x^B,
\end{aligned}\\
&C^{\mathfrak{n}} = - \frac{\tilde{N}}{2} k_2^{\mathfrak{n}}Y^{m}_{\ell}F_{n},\\
&B^{\mathfrak{n}} = \frac{p_b^2}{L_0^2} k_4^{\mathfrak{n}}Y^{m}_{\ell}F_{n}\text{d}\zeta - \left(k_1^{\mathfrak{n}}X^{m}_{\ell\;A} - |p_c|k_3^{\mathfrak{n}}Z^{m}_{\ell\;A}\right)
F_{n}\text{d}x^A.
\end{aligned}
\end{equation}

At second order in perturbations, the gravitational Hilbert-Einstein action (modulo hypersurface terms) shows a background-dependent part and a quadratic perturbative part. There are no linear terms in the perturbations, since the integration over all the $\chi$-constant sections retains only the zero-mode contribution, and the zero-modes are treated exactly in our formulation, removed from the mode expansions of the perturbations. The background part of the action reproduces the Legendre term corresponding to the Poisson-bracket structure shown in Eq. \eqref{backgroundbrac} together with a contribution to the total Hamiltonian constraint, which is given just by the background Hamiltonian \eqref{backgroundHam}. The perturbative part of the truncated action, quadratic in perturbations, has the form
\begin{equation}
\label{perturbationaction}
\int \left(\sum_{I=1}^6 \sum_{\mathfrak{n}} \frac{1}{\kappa} p_I^{\mathfrak{n}}\text{d}h_I^{\mathfrak{n}} - \left[\sum_{I=1}^4 C_I[k_I] + \tilde{H}_P[\tilde{N}]\right]\text{d}\chi \right),
\end{equation}
where the $C_I$'s are four perturbative constraints, arising from the linearization of the spacetime diffeomorphisms constraints of general relativity. The notation $C_I[k_I]$ indicates that $k_I^{\mathfrak{n}}$ are their perturbative Lagrange multipliers in our mode expansion, namely $C_I[k_I]=\sum C_I^{\mathfrak{n}}k_I^{\mathfrak{n}}$, as one would expect from the fact that $k_I^{\mathfrak{n}}$ are the mode coefficients that determine the perturbations of the lapse and shift. On the other hand, $\tilde{H}_{P}$ represents the perturbative contribution to the total Hamiltonian constraint. It is a sum of individual mode Hamiltonians. The total Hamiltonian constraint $\tilde{H}_T$ is a single and global constraint, obtained by integration over the entire $\chi$-constant section, and it is given by the sum of its background and perturbative parts: $\tilde{H}_T[\tilde{N}]=\tilde{H}_{KS}[\tilde{N}]+\tilde{H}_P[\tilde{N}]$. It is worth remarking that the associated Lagrange multiplier, namely the lapse $\tilde{N}$, does not vary on each $\chi$-constant section, so that indeed one only gets a constraint on the perturbed system. Moreover, the constraint is on the combined system formed by the background and its perturbations. In this sense, the constraint contains the backreaction of the perturbations on the background zero modes up to the quadratic perturbative order adopted in our truncation of the gravitational action. In other words, our Hamiltonian formalism is capable of retaining backreaction up to the aforementioned order. Finally, let us note that, from the above form of the perturbative contributions to the action, it is straightforward to realize that the axial and polar perturbations are decoupled in the Hamiltonian system, and their description naturally splits into separate contributions for each sector \cite{CPM1,JMMG1,CPM2}. 

The exact expressions of the perturbative gauge constraints $C_I$ and of the perturbative Hamiltonian $\tilde{H}_P$ are given in Appendix \ref{AppA}.

To relate the description of the perturbations of the spacetime metric corresponding to our foliation in terms of $\chi$-constant sections with the description associated to a foliation with $\zeta$-constant sections (which in the exterior geometry would reproduce the usual foliation in spatial sections), let us call $\,^{(ax)}g$ and $\,^{(po)}g$ the contributions of the axial and polar perturbations, respectively, to the metric two-form of the four-dimensional spacetime.  After a careful calculation, we then find that
\begin{equation}
\label{perturbation4metric}
\begin{aligned}
&\,^{(ax)}g^{m}_{\ell} = -\sum_n F_n \left[ 2\left( h_{1}^{\mathfrak{n}} X^{m}_{\ell\;A} \text{d}\zeta + \kappa k_{1}^{\mathfrak{n}} X^{m}_{\ell\;A} \text{d}\chi \right) \text{d}x^A - h_{2}^{\mathfrak{n}} X^{m}_{\ell\;AB} \text{d}x^A \text{d}x^B\right],\\
&\,^{(po)}g^{m}_{\ell} = \sum_n F_n \Bigg[ \left( h_{6}^{\mathfrak{n}} \text{d}\zeta^2 + 2 \kappa \frac{p_b^2}{L_0^2} k_{4}^{\mathfrak{n}}\text{d}\zeta \text{d}\chi - \kappa |p_c| \tilde{N} k_2^{\mathfrak{n}}\text{d}\chi^2 \right) Y^{m}_{\ell} \\
&+ 2 \left( h_5^{\mathfrak{n}}\text{d}\zeta + \kappa |p_c| k_{3}^{\mathfrak{n}}\text{d}\chi \right) Z^{m}_{\ell\;A}\text{d}x^A +  \left( h_3^{\mathfrak{n}}Y^{m}_{\ell\;AB} + h_4^{\mathfrak{n}}Z^{m}_{\ell\;AB} \right) \text{d}x^A \text{d}x^B \Bigg] .
\end{aligned}
\end{equation}
From these equations, it is easy to find the correspondence between the standard harmonic components of the metric perturbations in a 3+1 decomposition associated with a foliation in the $\zeta$-direction and our alternative 3+1 decomposition, which arises much more naturally in the interior region.

\section{Perturbative gauge invariants \label{SecIV}}

The fact that the perturbative gauge constraints are linear in the perturbations, and that they generate the perturbative diffeomorphisms that spoil the identification of which perturbative degrees of freedom have physical content and which are gauge, leads us to suggest a possible line of action. The idea is to change our perturbative variables so that the perturbative constraints are included directly as part of them. If we can do this while maintaining the use of canonical pairs, the gauge freedom will be directly restricted to the sector of the perturbative phase space where the gauge constraints reside. 

This proposal has two obstacles. First, the perturbative gauge constraints also contain background dependence. We can elude this complication by initially treating the background as fixed and restoring afterwards the canonical structure of the entire system, including the background, as we will explain below. The other complication is that the perturbative gauge constraints in principle do not commute under Poisson brackets, even if the background is kept fixed, preventing us from directly taking the four gauge constraints as new phase-space variables, each of them with a different canonically conjugate pair. However, since the perturbative gauge constraints arise from the linearization of the diffeomorphisms constraints of general relativity, it is possible to prove that one can always Abelianize them. That is, one can modify their expressions with a term that is linear in the perturbations and such that we obtain new constraints which commute at our order of perturbative truncation in the action. 

Concretely, this can be done by adding to the gauge constraint $C_2[k_2]$ of the polar sector a term proportional to the Hamiltonian background $\tilde{H}_{KS}$, multiplied by a factor that is linear in the perturbations. This term must be compensated with its negative in the truncated action; then, recalling that the Lagrange multipliers of the perturbative gauge constraints are also linear in the perturbations, we obtain a contribution in which the background Hamiltonian appears multiplied by a quadratic perturbative factor. This factor can be absorbed by a redefinition of the lapse $\tilde{N}$, which is the Lagrange multiplier of the total Hamiltonian constraint. In principle, this redefinition would also require an additional contribution in which such quadratic term multiplies the perturbative part of the total Hamiltonian constraint. However, this perturbative part is already quadratic in perturbations, so that its product by the modification to the lapse would be quartic, and therefore totally negligible in our perturbative truncation order. In the rest of our discussion, we assume this Abelianization of the perturbative gauge constraints and the redefinition of the lapse function $\tilde{N}$. The modes of the new polar gauge constraint replacing $C_2[k_2]$, which we call $\tilde{C}_2^{\mathfrak{n}}$, are given in Appendix \ref{AppA}. For simplicity, we do not change the notation employed for the lapse nor give the explicit expressions of its changes, which can be consulted in Refs. \cite{MM,MM2}.  

Hence, regarding the background as fixed, we can perform a linear canonical transformation in the perturbations so that they are described by gauge invariant canonical pairs, by the (Abelianized) perturbative constraints, and by variables conjugate to these constraints. We use the following notation for this transformation:
\begin{equation}
\label{GItransform}
\{h_I^{\mathfrak{n}},p_I^{\mathfrak{n}}\}_{I=1,...6}\longrightarrow \{Q_I^{\mathfrak{n}}, P_I^{\mathfrak{n}}\}_{I=1,...6} 
\end{equation}
such that
\begin{equation}
\label{momentaconstraints}
P_2^{\mathfrak{n}}= C_1^{\mathfrak{n}}, \quad P_4^{\mathfrak{n}}= \tilde{C}_2^{\mathfrak{n}}, \quad P_5^{\mathfrak{n}}= C_3^{\mathfrak{n}}, \quad P_6^{\mathfrak{n}}= C_4^{\mathfrak{n}}.
\end{equation}
Here, $C_I^{\mathfrak{n}}$, with $I=1,3,4$, are the mode coefficients of the corresponding perturbative gauge constraints introduced in Eq. \eqref{perturbationaction}. The change of perturbative variables given by Eq. \eqref{GItransform} can be considered a linear canonical transformation if the background is kept fixed, ignoring its dynamical nature. An explicit canonical transformation of this type was given in Refs. \cite{MM,MM2}. This canonical transformation can be treated as a composition of two independent ones, one for the axial sector and the other for the polar sector. The important point that we want to stress here is that the perturbative gauge constraints can be chosen in this way as part of a new canonical set for the perturbations. This is facilitated by the fact that such gauge constraints are linear in the original perturbative variables. We then conclude that their canonically conjugate variables are pure gauge, with values that are changed in the gauge orbits, and therefore do not correspond to genuine degrees of freedom. The subset of variables formed by the gauge constraints and their canonical pairs simply reflects the freedom to perform perturbative diffeomorphisms changing the identification of points between the original and the perturbed manifold and how this freedom can be parametrized in the phase space of the perturbations. The perturbative quantities that are not affected by these changes are the perturbative gauge invariants. In our Hamiltonian formalism, determining all such possible gauge invariants is immediate. Modulo perturbative gauge constraints, they can be identified with the functions (linear functions if we restrict our discussion to linear perturbations) of the two canonical pairs that commute with those constraints, so that they do remain invariant under the transformations generated by perturbative diffeomorphisms. With our notation, this means that $\{Q_1^{\mathfrak{n}},P_1^{\mathfrak{n}},Q_3^{\mathfrak{n}},P_3^{\mathfrak{n}}\}$ form a complete, canonical set of variables on the gauge invariant sector of the phase space of the perturbations, with the first pair describing the gauge invariant axial perturbations and the second pair the polar part. The expression of this set of perturbative gauge invariants is given in Appendix \ref{AppA}.

In the previous paragraph, we have commented that our transformation is canonical only if we do not treat the background as a dynamical entity. However, it is possible to show that the transformation can actually be extended to a canonical one on the whole phase space of the combined system formed by the perturbations and the background, at least at the perturbative order of our truncation in the action. Since this truncation is quadratic in the perturbations, it includes the effect of perturbative quadratic modifications to the zero modes, in the same way as the total Hamiltonian constraint acquires a contribution that is quadratic in the perturbations. Taking this into account, it is possible to prove that the background variables can always be modified with terms that are quadratic in the perturbations and that restore the canonical structure in the total phase space that describes background and perturbations. The form of this modification of the background variables is general once we know the part of the canonical transformation for the perturbative variables, and is given in Appendix \ref{AppB}. The modification of the zero modes can then be computed at the end of all the canonical perturbations that we may implement on the perturbative sector. In order to maintain our notation as simple as possible, we will not change the notation for our background variables, using the same symbols for the original and the modified variables. The difference can be easily inferred from the context at each step of our analysis. 

The modification of the background variables has however an important consequence. The change in the Legendre terms to arrive at new canonical zero modes leads to new quadratic perturbative contributions to the total Hamiltonian constraint as a byproduct. Although these new contributions to the Hamiltonian can be deduced explicitly using the formulas for the corrections to the zero modes \cite{LMM}, there is a short cut in the derivation that consistently produces the same result. It suffices to realize that the change in the Hamiltonian constraint comes from the fact that the canonical transformation on the perturbative sector is background dependent and, since this background is dynamical, it changes the dynamics by assigning part of it to the background rather than to the perturbations. To account for this, one only needs to recall that the Hamiltonian changes when one implements an explicitly time-dependent canonical transformation, adding the derivative of the generator of the transformation with respect to such explicit time dependence. In our Hamiltonian formalism, this is the derivative in the dynamical $\chi$-direction, acting on the background dependence of the generator of the canonical transformation of the perturbations. But, for the background, such dynamical derivative is simply given by the Poisson bracket with the background part of the total Hamiltonian constraint, $\tilde{H}_{KS}[\tilde{N}]$. In this way, it is extremely simple to appropriately correct the perturbative contribution to the total Hamiltonian constraint when performing a canonical change of perturbative variables without paying special attention to the associated modification of the zero modes that is necessary to retain the global canonical structure in our system. In the following, we will adopt this procedure without saying it explicitly. Moreover, as far as the description of the dynamics of the perturbations is concerned, and since the corresponding mode equations are linear in the perturbations, as it corresponds to a quadratic Hamiltonian, it is totally justified to ignore corrections of the background that are at least quadratic in the perturbations. In particular, this allows us to use the background Hamiltonian to simplify the expression of the perturbative contribution to the total Hamiltonian constraint, because such constraint implies that the background Hamiltonian is at most of quadratic order in our perturbative hierarchy, and changes of this order in the already quadratic perturbative part of the constraint are negligible in our truncation.

Finally, after performing a canonical transformation of the type discussed above, the resulting expression for the perturbative Hamiltonian typically contains contributions proportional to the perturbative gauge constraints or to the background Hamiltonian. In the former case, the factors multiplying the perturbative constraints are necessarily linear in perturbations, since the constraints themselves are linear and the Hamiltonian is retained only up to quadratic order in our perturbative truncation of the action. In the latter case, namely for terms proportional to the background Hamiltonian, the accompanying factors that appear are quadratic in the perturbations for similar reasons. Both cases lead to a redefinition of Lagrange multipliers in the truncated action of the system. For the Lagrange multipliers of the perturbative gauge constraints, the redefinition absorbs linear perturbative corrections, as we have explained, while for the case of the background Hamiltonian, we can absorb the quadratic perturbative factors in a redefinition of the lapse $\tilde{N}$, in a similar way as we did in our Abelianization of the perturbative gauge constraints. In the rest of our discussion, we implicitly assume that all the Lagrange multipliers have been properly redefined without changing the notation employed for them, nor giving the explicit expressions for their changes, which can be consulted in Refs. \cite{MM,MM2}.  

With all these ingredients and after a careful calculation, the perturbative part of the resulting, total Hamiltonian constraint can be shown to depend quadratically on the perturbations only through the perturbative gauge invariants, as it was expected by consistency, since this constraint must itself be gauge invariant. Hence, it has the form 
\begin{equation}
\label{cuadraHami}
\mathbf{\tilde{H}}[\tilde{N}] =  \sum_{\mathfrak{n}}\frac{\tilde{N}}{2\kappa}\left[A_{(ax)}\,(Q_1^{\mathfrak{n}})^2 + B_{(ax)}\,(P_1^{\mathfrak{n}})^2 + C_{(ax)}\,Q_1^{\mathfrak{n}}P_1^{\mathfrak{n}} + A_{(po)}\,(Q_3^{\mathfrak{n}})^2 + B_{(po)}\,(P_3^{\mathfrak{n}})^2 + C_{(po)}\,Q_3^{\mathfrak{n}}P_3^{\mathfrak{n}}\right].
\end{equation}
The expression of the new Hamiltonian coefficients are given in Appendix \ref{AppA}. Note that this perturbative contribution to the total Hamiltonian constraint still splits into an axial and a polar part, and each of them in different modes, all decoupled from one another. 

\section{Master functions \label{SecV}}

In order to establish a more direct connection with the usual description of perturbative gauge invariants in terms of the master functions employed in BH physics, we discuss in this section how to select some preferred set of perturbative gauge invariants among all those that can be reached by means of background-dependent linear combinations of the canonical pairs that we have selected for the modes of the axial and polar physical degrees of freedom. More specifically, we want to find the mode variables corresponding to the Cunningham-Price-Moncrief (CPM) \cite{CPM1} and Zerilli-Moncrief (ZM) \cite{Zer,Mon} master functions for the axial and polar perturbations, respectively. 

The form of these gauge invariants can be derived starting with their familiar expressions in terms of metric perturbations, particularizing them to our 3+1 decomposition of the metric using Eq. \eqref{perturbation4metric}, and rewriting everything in our notation for the metric background functions and with our changes of canonical perturbative variables. In doing this, one can also associate the resulting mode variables with canonically conjugate pairs for them. This lengthy analysis leads to the expressions that we give below (see also Refs. \cite{MGAC,MGA}).  

Starting with the perturbative gauge invariants introduced in the previous section and specified in Appendix \ref{AppA}, the mode variables corresponding to the CPM master function are \cite{MGAC}
\begin{equation}
\label{CPM}
\begin{aligned}
& \mathcal{Q}_{CPM}^{\mathfrak{n}} = -\sqrt{\frac{(\ell-2)!}{(\ell+2)!}|p_c|} \bigg[Q_1^{\mathfrak{n}} + 4\ell(\ell+1)\frac{L_0^2 \Omega_b}{p_b^2}P_1^{\mathfrak{n}}\bigg]  , \\ 
& \mathcal{P}_{CPM}^{\mathfrak{n}} = -\sqrt{\frac{(\ell+2)!}{(\ell-2)!}\frac{1}{|p_c|}} \bigg[P_1^{\mathfrak{n}} - \frac{(\ell-2)!}{(\ell+2)!}\Omega_b\bigg(Q_1^{\mathfrak{n}} + 4\ell(\ell+1)\frac{L_0^2 \Omega_b}{p_b^2}P_1^{\mathfrak{n}}\bigg)\bigg] ,
\end{aligned}
\end{equation} 
and, for the ZM master function \cite{MGA},
\begin{equation}
\label{ZM}
\begin{aligned}
& \mathcal{Q}_{ZM}^{\mathfrak{n}} = \sqrt{\frac{(\ell+2)(\ell-1)}{\ell(\ell+1)}|p_c|}Q_3^{\mathfrak{n}}, \\ & \mathcal{P}_{ZM}^{\mathfrak{n}} = \sqrt{\frac{\ell(\ell+1)}{(\ell+2)(\ell-1)}\frac{1}{|p_c|}}\bigg[P_3^{\mathfrak{n}} + \frac{(\ell+2)(\ell-1)}{2\ell(\ell+1)} \bigg(C_{(po)} + 2\Omega_b\bigg) Q_3^{\mathfrak{n}}\bigg] .
\end{aligned}
\end{equation}
Here, $C_{(po)}$ is the background-dependent coefficient of the polar perturbative Hamiltonian \eqref{cuadraHami}, and can be found in Appendix \ref{AppA}. 

Strictly speaking, the above configuration variables are not exactly the mode coefficients of the CPM and ZM master functions, but differ from them only by global constant factors that depend on $\ell$, as discussed in Ref. \cite{MGA}: such mode coefficients are $\mathcal{Q}_{CPM}^{\mathfrak{n}}$ and $\mathcal{Q}_{ZM}^{\mathfrak{n}} $ respectively multiplied by $2\sqrt{(\ell-2)!/(\ell+2)!}$ and $2\sqrt{(\ell+2)(\ell-1)/[\ell(\ell+1)]^{3}}$. These factors are not included here because omitting them results in a more compact and manageable Hamiltonian structure, as will become apparent below. Nevertheless, since the CPM and ZM master functions can be easily retrieved from these configuration variables by a trivial operation whithout any dynamical content, we will refer to them as the CPM and ZM variables, in a harmless abuse of the language.

After performing the background-dependent canonical transformation of the perturbations that leads to these new mode variables, the perturbative Hamiltonian takes a simple quadratic form, with no crossed term between configuration and momentum variables, and with the coefficients of the squared contributions of the momenta all equal to one. This last fact immediately implies that the Hamiltonian derivative of the configuration gauge invariants is proportional to their respective canonical momenta. Moreover, the new perturbative Hamiltonian is \cite{MGAC,MGA}
\begin{equation}
\label{masterHamiltonian}
\tilde{\mathcal{H}}[\tilde{N}] = \sum_{\mathfrak{n}} \frac{\tilde{N}}{2\kappa} |p_c| \left[ \left(\mathcal{P}_{CPM}^{\mathfrak{n}}\right)^2 + \left(\omega_n^2 - V_{\ell}^{RW}\right) \left(\mathcal{Q}_{CPM}^{\mathfrak{n}}\right)^2 + \left(\mathcal{P}_{ZM}^{\mathfrak{n}}\right)^2 + \left(\omega_n^2 - V_{\ell}^{Z}\right) \left(\mathcal{Q}_{ZM}^{\mathfrak{n}}\right)^2 \right],
\end{equation}
where the Fourier frequency $\omega_n$ appears only in the quadratic terms in the configuration variables and is completely isolated from other contributions, and $V_{\ell}^{RW}$ and $V_{\ell}^{Z}$ are the so-called Regge-Wheeler (RW) \cite{RW} and Zerilli (Z) \cite{Zer} potentials, expressed in terms of our background variables, namely
\begin{equation}
\label{masterpotentials}
\begin{aligned}
&V_{\ell}^{RW} = -\frac{1}{p_c^2} \left[ \ell(\ell+1)\frac{p_b^2}{L_0^2} + 6\Omega_b \Omega_c \right],\\
&V_{\ell}^{Z} = -\frac{1}{p_c^2}\left[(\ell+2)^2(\ell-1)^2\left(\ell(\ell+1)\frac{p_b^2}{L_0^2} - 6\Omega_b\Omega_c\right) + \left(6\Omega_b\Omega_c\frac{L_0^2}{p_b^2}\right)^2\left((\ell+2)(\ell-1)\frac{p_b^2}{L_0^2} - 2\Omega_b\Omega_c\right)\right]\frac{1}{\Lambda_{\ell}^2}.
\end{aligned}
\end{equation}
In this equation, we have introduced the notation $\Lambda_{\ell} = (\ell+2)(\ell-1) - 6L_0^2\Omega_b\Omega_c /p_b^2$.  

We note that, with our transformation, a global factor of $|p_c|$ arises in the perturbative Hamiltonian. This factor suggests a change in the natural evolution parameter to describe the dynamics of the perturbations. In direct analogy with the use of tortoise coordinates in the Schwarzschild solutions, one can introduce the reparametrization $d\eta = |p_c|\tilde{N}d\chi$, so that the Hamiltonian \eqref{masterHamiltonian} with the prefactor $\tilde{N}|p_c|$ removed generates evolution along the $\eta$-direction for the perturbations. Actually, on the Schwarzschild solutions, we get $d\eta=(1-2M/\chi)^{-1} d\chi$, so that one recovers the standard tortoise coordinate. In the following, we denote the derivative along this $\eta$-direction indistinctly as $\partial_{\eta}$ or by a prime. 

We can sum over all the Fourier modes to pass from dynamical mode variables for the gauge invariant perturbations to fields in the two-dimensional manifold that describes the set of spherical orbits in our background. We obtain an axial and a polar field for each pair of labels of the spherical harmonics $(\ell, m)$. So, we retain these labels also for the corresponding fields in two dimensions. The corresponding configuration fields are, strictly speaking, what we usually call master functions. The perturbative axial and polar Hamiltonians that we have obtained above can be easily seen to lead then to wave equations on this two-dimensional manifold for the CPM and ZM master functions, which have the following form:
\begin{equation}
\label{2wave}
    \left[ -\partial_{\zeta}^2 + \partial_{\eta}^2 - V_{\ell} \right] \mathcal{Q}_{\ell}^{m} = 0, 
\end{equation}
where we have omitted the label $CPM$ or $ZM$ in the master function $\mathcal{Q}_{\ell}^{m}$, and the label $RW$ or $Z$ in the corresponding potential. Up to a global factor, the above expression can also be written in terms of the two-dimensional Laplace-Beltrami operator associated with the background metric induced on the considered two-dimensional manifold of different spherical orbits.

As an enlightening exercise, we can evaluate our expressions on the Schwarzschild solutions, using the relations \eqref{Schwarz}, which hold on shell in this case, and neglecting quadratic perturbative corrections in the field equations of our gauge invariants. A simple calculation shows that one recovers then the standard expressions of the RW and Z potentials \cite{MGAC,MGA}:
\begin{equation}
\label{Schwarzpotentials}
\begin{aligned}
&\,^{(Sc)}V_{\ell}^{RW} = \left(1-\frac{2M}{\chi}\right)\frac{1}{\chi^2}\left[\ell(\ell+1) - 3\frac{2M}{\chi}\right],\\ 
&\,^{(Sc)}V_{\ell}^{Z} = \left(1-\frac{2M}{\chi}\right)\frac{1}{\chi^2}\left[(\ell+2)^2(\ell-1)^2\left(\ell(\ell+1) + 3\frac{2M}{\chi}\right) + \frac{36M^2}{\chi^2}\left((\ell+2)(\ell-1) + \frac{2M}{\chi}\right)\right]\frac{1}{\,^{(Sc)}\Lambda_{\ell}^2 },
\end{aligned}
\end{equation}
with the label $(Sc)$ indicating evaluation on Schwarzschild solutions and $\,^{(Sc)}\Lambda_{\ell} = (\ell+2)(\ell-1) + 6M/\chi$. In addition, the definitions of the CPM and ZM master functions introduced above can be rewritten in terms of metric perturbations. These perturbations can in turn be expressed through the components of the metric in a 3+1 decomposition corresponding to a foliation in $\zeta$-constant sections, by making use of the relations \eqref{perturbation4metric}, the identities \eqref{Schwarz} for the Schwarzschild solutions, the on-shell equations of general relativity that these solutions satisfy, and taking into account the previous comments on the $\ell$-dependent global constant factors. In this way one can check in full detail that the aforementioned master functions admit their familiar expression in terms of metric perturbations in the case of Schwarzschild \cite{MGAC}. In this sense, our definitions are a generalization of these expressions, valid for any background, and without on-shell restrictions. For instance, this opens the possibility of considering backgrounds that are corrected with respect to their Einsteinian analogs by effective modifications, even of quantum nature, and employ our definition of gauge invariants and master functions for perturbations around such backgrounds.  

Whereas the privileged role that the CPM and ZM master functions can be understood in our formalism as a direct consequence of the simplicity of their associated Hamiltonian, which is diagonal, with constant square-momentum coefficient, and with an isolated contribution of the (squared) Fourier frequency, the question remains on whether these Hamiltonian features exhaust the freedom in the choice of master functions, or there are other admissible choices with the same properties. In the next section we show that in fact there exists a whole infinite family of canonical pairs for the gauge invariant perturbations with the desired Hamiltonian structure. The relation between different members in this family is given by background-dependent canonical transformations, which are in direct relation with the Darboux transformations that we succinctly described in the Introduction. 

\section{Darboux transformations \label{SecVI}}

In the previous section, we determined the perturbative mode variables corresponding to the CPM and ZM master functions and to perturbative momenta canonically conjugate to them. We showed that the perturbative contribution to the total Hamiltonian corresponding to these master functions and their canonical momenta has a very specific structure. In terms of the evolution coordinate $\eta$, the perturbative Hamiltonian has the form of a sum over two infinite collections of generalized harmonic oscillators, one for the axial sector and another one for the polar sector, with an isolated contribution of the squared Fourier frequency and an $\ell$-dependent and background-dependent potential, which is given by an off-shell generalization of the RW or the Z potential (respectively for the axial and polar sectors). The axial and polar perturbative Hamiltonians lead then to dynamical equations that certainly reproduce the CPM and ZM master equations on shell, but in addition generalize them off shell to backgrounds that in principle need not satisfy the relativistic equations.     

In general relativity, it is known that the Darboux transformations relate the CPM and ZM master equations to similar equations but with other different potentials. We recall that, starting from one of our master functions $\mathcal{Q}_{\ell}^{m}$, if we introduce a new perturbative gauge invariant $\tilde{\mathcal{Q}}_{\ell}^{m}$ such that $\tilde{\mathcal{Q}}_{\ell}^{m} \propto (\mathcal{Q}_{\ell}^{m})^{\prime}+ g_{\ell} \mathcal{Q}_{\ell}^{m}$, where $g_{\ell}$ is a solution to the Riccati equation $g_{\ell}^{\prime}-g_{\ell}^2+V_{\ell}+c_{\ell}=0$ with $c_{\ell}$ being a constant, then $\tilde{\mathcal{Q}}_{\ell}^{m}$ provides a new master function that satisfies again a generalized wave equation in which the original potential $V_{\ell}$ is replaced with $v_{\ell}$. Moreover, the new and the original potentials are related by $v_{\ell}=V_{\ell}+2g_{\ell}^{\prime}$.

In this section, we are going to show that Darboux transformations are in one-to-one correspondence with canonical transformations on the perturbative gauge invariants that respect the structure of the Hamiltonian found above. With this aim, let us consider a generic canonical transformation of our perturbative mode variables that do not mix decoupled modes. For the moment being, we will not even consider the possibility of violating polarity. Ignoring the $CPM$ or $ZM$ ($RW$ or $Z$, respectively) labels, we then consider linear background-dependent transformations of the form \cite{MGAC}
\begin{equation}
\label{canoperturb}
\begin{aligned}
&\mathcal{Q}^{\mathfrak{n}} = A_{\mathfrak{n}}\tilde{\mathcal{Q}}^{\mathfrak{n}}+ B_{\mathfrak{n}} \tilde{\mathcal{P}}^{\mathfrak{n}},\\
&\mathcal{P}^{\mathfrak{n}} = C_{\mathfrak{n}}\tilde{\mathcal{Q}}^{\mathfrak{n}}+ D_{\mathfrak{n}} \tilde{\mathcal{P}}^{\mathfrak{n}}.
\end{aligned}
\end{equation}
The new canonical mode variables are distinguished with a tilde-notation. The background-dependent coefficients $A_{\mathfrak{n}}$, $B_{\mathfrak{n}}$, $C_{\mathfrak{n}}$, and $D_{\mathfrak{n}}$
are subject to the condition that the transformation is canonical, which for each value of $\mathfrak{n}$
implies 
\begin{equation}
\label{symplectic}
A_{\mathfrak{n}}D_{\mathfrak{n}}-B_{\mathfrak{n}}C_{\mathfrak{n}}=1.
\end{equation}

We assume that $B_{\mathfrak{n}}$ is not vanishing for all $\mathfrak{n}$, so that the transformation is not just a simple redefinition of the gauge invariant modes, not even for some of them. Furthermore, to remove the freedom in background-dependent scalings, we can always make $B_{\mathfrak{n}}$ a nonzero constant.

Owing to the change of perturbative variables and to the background dependence of the transformation, we get both a new functional form for the original perturbative Hamiltonian and new perturbative contributions to this Hamiltonian. Using the techniques that we have already explained, it is not difficult to derive the expression of the new perturbative Hamiltonian \cite{MGAC}. To arrive at a Hamiltonian with the same structure as before, and adopting a dynamical description in the $\eta$-direction, we require again the crossed terms between perturbative configuration and momentum variables to vanish, and the coefficient of the squared perturbative momentum variables to be one half. Together with the constancy of $B_{\mathfrak{n}}$, the canonical condition \eqref{symplectic} and the absence of crossed terms in the Hamiltonian lead to \cite{MGAC}
\begin{equation}
\label{cross}
D_{\mathfrak{n}}=A_{\mathfrak{n}}, \quad C_{\mathfrak{n}}=\frac{A_{\mathfrak{n}}^2-1}{B_{\mathfrak{n}}}.
\end{equation}
We see that, as expected, the condition that the perturbative Hamiltonian contains no crossed terms allows us to fix the choice of the new perturbative momenta in terms of only the definition of the new perturbative configuration variables. As for the latter, remarkably, calling $g_{\mathfrak{n}}=  -A_{\mathfrak{n}}/B_{\mathfrak{n}}$, the demand that the coefficient of the term with the squared momentum be one half amounts then to a Riccati equation of the form \cite{MGAC}
\begin{equation}
\label{Riccati}
g_{\mathfrak{n}}^{\prime}-g_{\mathfrak{n}}^2+\left(V_{\ell} - \omega_{n}^2\right)= -\frac{1}{B_{\mathfrak{n}}^2}.
\end{equation} 
If the coefficients of our canonical transformation vary on the $\eta$-direction owing exclusively to their dependence on the background zero modes (that is, if they do not display any explicit $\eta$-dependence, but only an implicit one through the background dynamics), the same applies to $g_{\mathfrak{n}}$. Then, employing the definition of $\eta$ and the fact that the $\chi$-evolution of the background is generated by the background Hamiltonian, we have that the $\eta$-derivative in the above equation is not other than the Poisson bracket of $g_{\mathfrak{n}}$ with $\tilde{H}_{KS}[\tilde{N}]$ (with respect to the background symplectic structure) divided by $|p_c|$. Therefore, the derivative can be computed in this case exclusively from the background phase space structures and without an explicit knowledge of the solutions to the background equations of motion.

Equation \eqref{Riccati} resembles to the Riccati equation associated to the Darboux transformations. The main difference is the dependence on the Fourier frequency $\omega_n$, which contributes with its square and potentially in the possible dependence on the Fourier label $n$ of both $g_{\mathfrak{n}}$ and $B_{\mathfrak{n}}$. A generic dependence on the Fourier frequency in our canonical transformation would introduce a nontrivial nonlocality, since we recall that finding $\omega_n$ requires spectral techniques. Furthermore, a dependence of this type would lead to a new potential in which the squared frequency contribution would not be isolated in the configuration part. In turn, this would prevent us from expressing the dynamical equations of the corresponding master function as a generalized wave equation in two dimensions, like in Eq. \eqref{2wave}. This very unpleasant physical features can be avoided only by requiring a specific dependence of $B_{\mathfrak{n}}$ on $\omega_n$, namely we must have $B_{\mathfrak{n}}^{-2}=\omega_n^2+c_{\ell}$. In this case, it is straightforward to see that Eq. \eqref{Riccati} reproduces the equation found in the Darboux transformations. 

In summary, taking $B_{\mathfrak{n}}$ of the above form and calling in this case $g_{\mathfrak{n}}=-A_{\mathfrak{n}}\sqrt{\omega_n^2+c_{\ell}}=g_{\ell}$, since it satisfies a differential equation that only depends on $\mathfrak{n}$ through $\ell$, we conclude that, in order for the coefficient of the squared momentum term in the perturbative Hamiltonian to be equal to one half, we must have $g_{\ell}^{\prime}-g_{\ell}^2+V_{\ell}+c_{\ell}=0$. For each possible constant value of $c_{\ell}$, this equation, which is precisely the same that characterizes the Darboux transformation, fully determines our canonical transformation. In particular, the relevant information about $g_{\ell}$ is contained, up to a suitable constant, in the coefficient relating the old and new configuration mode variables in the transformation \eqref{canoperturb}. 

Gathering all the information about the transformation of the perturbative mode variables, we conclude that the new variables must have the following appearance, given by the inverse of the canonical transformation \eqref{canoperturb}:
\begin{equation}
\label{Darbouxform}
\begin{aligned}
&\tilde{\mathcal{Q}}^{\mathfrak{n}}= -\frac{g_{\ell}}{\sqrt{\omega_n^2+c_{\ell}}} \mathcal{Q}^{\mathfrak{n}}- \frac{1}{\sqrt{\omega_n^2+c_{\ell}}}\mathcal{P}^{\mathfrak{n}},\\
&\tilde{\mathcal{P}}^{\mathfrak{n}}=\left(\sqrt{\omega_n^2+c_{\ell}} - \frac{g_{\ell}^2}{\sqrt{\omega_n^2+c_{\ell}}}\right)\mathcal{Q}^{\mathfrak{n}}- \frac{g_{\ell}}{\sqrt{\omega_n^2+c_{\ell}}}\mathcal{P}^{\mathfrak{n}}.
\end{aligned}
\end{equation} 
Upon substituting the Hamiltonian equation of motion for the momenta, this clearly shows that $\tilde{\mathcal{Q}}^{\mathfrak{n}} \propto (\mathcal{Q}^{\mathfrak{n}})^{\prime}+ g_{\ell} \mathcal{Q}^{\mathfrak{n}}$. Moreover, each canonical transformation in the analyzed class is totally characterized by a constant $c_{\ell}$ (independent of the Fourier frequency) and a solution to the Riccati equation \eqref{Riccati}, exactly as it happens with the Darboux transformations that do not mix polarity. Realizing also that our canonical transformations lead to generalized wave equations in two dimensions for the resulting master functions associated to the perturbative configuration variables, which are exactly of the same form as those reached with the polarity preserving Darboux transformations, we have proven that there is indeed a one-to-one correspondence between the two different types of transformations: the canonical ones and the Darboux ones. We also notice that the composition of two canonical transformation respecting the structure of the perturbative Hamiltonian is itself a canonical transformation with such properties, in the same way as a composition of Darboux transformations is a new Darboux transformation (in both cases, respecting polarity at this stage of our discussion). Hence, in both situations we find a group of transformations, which clearly includes the identity.

It is worth remarking that the above analysis can be carried out independently of the form of the original potential $V_{\ell}$. Hence, it is equally valid in the case of the axial CPM master function with the RW potential as in the case of the polar ZM master function with the Z potential. In conclusion, in both cases we find a direct relation between the canonical transformations that preserve the form of the corresponding perturbative Hamiltonian (up to a change in the potential) and the Darboux transformations between master functions of that polarity sector. 

It is illustrative to compare the different steps that lead in the Lagrangian and Hamiltonian formalisms to the two types of transformations that we are considering in our nonrotating BH scenarios, and which we have related by a bijective correspondence. In the Lagrangian formalism \cite{CM1,Lenzi:2021wpc,CM2}, a) one first has to find the gauge invariant combinations of the metric perturbations to isolate the physical degrees of freedom of the perturbations. b) Next, one constructs master functions as combinations of gauge invariant perturbations and their first derivatives in the non-Killing direction of the two-dimensional metric induced on the set of (spherical) symmetry orbits. These master functions satisfy generalized wave equations obtained from the Laplace-Beltrami operator associated to the aforementioned two-dimensional metric. c) Finally, Darboux transformations relate these master functions, changing the potential to which they are subject in the generalized wave equation. In our Hamiltonian formalism, on the other hand, a) one first introduces a canonical transformation to perturbative variables that commute with the perturbative constraints. These variables can be understood as phase space coordinates for the sector of gauge invariant perturbations. b) Next, a second canonical transformation in this gauge invariant sector leads to new perturbative variables in which the configuration part can be understood as the mode coefficients of master functions that satisfy generalized wave equations in two dimensions (as in the standard Lagrangian description). We have implemented this second transformation for a specific, standard choice of the master functions, but we have seen that other choices are available, all of them with a characteristic form of the perturbative Hamiltonian, in which only the potential changes. c) Finally, we have seen that all these choices are related in fact by a specific family of canonical transformations.    

Let us finish our discussion by considering the case of transformations that violate parity. For this, it is particularly convenient to first perform a transformation in the polar sector that changes the ZM master function and its Z potential to the master function with the same RW potential as in the axial sector. This transformation is explicitly given by
\begin{equation}
g_{\ell} = \frac{6L_0^2\Omega_b^2\Omega_c}{p_b^2|p_c|\Lambda_{\ell}} - \frac{1}{|p_c|}\frac{(\ell+2)!}{(\ell-2)!}\bigg(\frac{p_b^2}{L_0^2}\frac{1}{\Omega_b}\bigg)^2\frac{1}{12\Omega_c},\qquad c_{\ell} =\bigg[\frac{1}{|p_c|}\frac{(\ell+2)!}{(\ell-2)!}\bigg(\frac{p_b^2}{L_0^2}\frac{1}{\Omega_b}\bigg)^2\frac{1}{12\Omega_c}\bigg]^2.
\end{equation}
Although it may not be apparent, $c_{\ell}$ remains constant under the evolution generated by the background Hamiltonian. In this way, and after performing a convenient redefinition of the lapse function, the two sectors are described by the same kind of master function, with perturbative Hamiltonians that formally coincide. In our Hamiltonian formalism, it is straightforward to see that, if we ignore polarity and regard in the following axial and polar mode variables (for the same value of $\mathfrak{n}$) as two indistinguishable generalized harmonic oscillators, any rotation in the two-dimensional configuration space corresponding to these oscillators, accompanied by the same rotation in the conjugate momentum space, is a canonical transformation that preserves the form of the combined perturbative Hamiltonian formed by the axial and polar contributions (adopting the $\eta$-direction as the evolution direction, as before). All such phase space rotations can then be considered Darboux transformations of the special kind that interchanges and mixes the polarity of the master functions. In particular, Chandrasekhar's transformation \cite{BHM} corresponds to the rotation by an angle of $\pi/2$ (combined with a flip of sign in one of the canonical pairs if one wants to preserve the sign of the configuration variables, and with the initial polar transformation from the Z to the RW potential if one wants to start with the ZM master function). 

\section{Conclusion \label{SecVII}}

In this work, we have reviewed a Hamiltonian formalism developed for perturbed nonrotating and uncharged BH spacetimes. The construction of this formalism starts from the geometry of the interior of the BH, which can be described as a KS cosmology, but the outcome can be naturally extended to account for the interior and exterior geometries. The extension is not provided by a generalized Wick rotation, but rather by a complex canonical transformation on the phase space of the background spacetime or, alternatively, by allowing extended ranges for some basic real functions on that phase space. All relevant quantities appearing in the analysis of the perturbations depend on the background exclusively via such functions. Lately, other Hamiltonian proposals for BH backgrounds that identify the interior timelike direction as the dynamical direction of evolution have appeared in the literature \cite{Eter,Uncertain}. As far as we know and to date, the works reviewed here provide the only complete Hamiltonian formulation that successfully includes not only the background, but also its perturbations, both with polar and axial polarities.

This Hamiltonian formalism and the canonical symplectic structure employed in it have been derived from the gravitational action of general relativity (ignoring hypersurface terms), truncating its perturbative expansion to quadratic order in the perturbations, which is the first subleading order in the expansion of the action. The techniques described here can in principle be applied as well to other gravitational actions different from the Hilbert-Einstein one, or including matter contributions instead of the vacuum case. It would be interesting to investigate these further applications, for example studying BH backgrounds arising in modified general relativity (see e.g. Refs. \cite{modif1,modif2,modif3}), or with dark matter halos \cite{halos}. A remarkable feature of our formalism, even without considering generalizations of the Hilbert-Einstein action, is that it is an off-shell construction, in which not even the perturbative gauge is fixed. This allows us to employ it almost straightforwardly in scenarios in which the deviations of the background with respect to general relativity are due to effective modifications, irrespective of their origin; for instance, we can consider the incorporation of effective corrections owing to quantum geometry phenomena \cite{AOS,AOS2}, or smearing processes, etc. 

The fact that we did not fix the perturbative gauge has been possible because in our Hamiltonian formalism it is a simple task to identify the generators of perturbative gauge transformations, obtained from the linearization of the diffeomorphism constraints of general relativity. Then, using that they are linear in first-order perturbations, we can complete them into a full canonical set of perturbative variables for our gravitational system. Canonical pairs that commute with these perturbative constraints are directly identified as good basic variables in the perturbative gauge invariant sector of the model. Any perturbative gauge invariant is a linear function(al) of them, with coefficients that are allowed to be background dependent. Notably, such a canonical set of perturbative variables, in which gauge invariants are immediately identifiable, can always be extended to include also canonical pairs for the background, corresponding to zero modes which contain quadratic corrections of the perturbations to preserve the global canonical structure. 
  
Furthermore, we have shown how the perturbative gauge invariants can be accommodated in functions on the two-dimensional set of symmetry orbits of the BH spacetime, $\Sigma_2$. These functions provide the master functions which are usually employed in BH perturbation theory. In our formalism, they appear together with canonically conjugated momentum fields. We have determined pairs such that the configuration fields are the CMP and ZM master functions for the axial and polar cases, respectively.  We have demonstrated that the field equations for these master functions are generalized wave equations, with potentials that generalize the RW and Z potentials, in the sense that they are expressed as functions on the background phase space, without the need to evaluate them on solutions, but reproducing the standard results if this evaluation is done, namely in the case of the Schwarzschild geometry for general relativity. In particular, our Hamiltonian formalism allows us to have at our disposal all the mathematical tools that are required for the calculation of quasinormal modes within an off-shell framework, which opens a window to the study of effective backgrounds. 

These effective descriptions may arise from quantum corrections, as we have commented. Furthermore, the fact that we have a complete Hamiltonian formulation of our gravitational system, containing the BH background and its geometric perturbations, in which we have a simple canonical symplectic structure, with fundamental variables that can easily been represented quantum mechanically, facilitates enormously the passage to a quantum version of the model. An appealing possibility is to adopt a representation, in the same spirit as in loop quantum cosmology \cite{LQC,ABRev}, of the holonomy-flux algebra that can be constructed from our tetrad and connection variables for the background (following techniques inspired by loop quantum gravity \cite{LQG,Thiemann,LQCvel}). This loop representation can be combined with a more standard, Fock representation of the mode coefficients that correspond to the perturbative master functions in a Fourier expansion \cite{MM,MM2,MGA}. This combination of representations is usually called hybrid representation in the literature about loop quantum cosmology \cite{hybrid}. The quantization is completed with a representation of the constraints \cite{MM,MM2}. The perturbative gauge constraints are easy to represent as generalized derivatives, and their imposition restricts physical states to depend exclusively on background and perturbative gauge invariants. The only nontrivial constraint is the total Hamiltonian constraint, formed by the background Hamiltonian and the quadratic perturbative Hamiltonians for the axial and polar gauge invariants. Nonetheless, even this task eventually turns out to be relatively straightforward to achieve \cite{MGA}, because the background dependence that appears in the perturbative Hamiltonians is simple (they essentially are the Hamiltonians associated with the RW and Z potentials). Moreover, this dependence can be further simplified by substituting the background Hamiltonian as it vanished, something that is legitimate at the truncation order of our formulation \cite{MGA}. The only step left in the quantization program is the resolution of this quantum constraint and the construction of physical observables acting on its solutions. In summary, our formalism provides a canonical system which is optimally prepared for its hybrid quantization, allowing the quantum description of perturbed nonrotating BH geometries and a direct road to discuss their quantum physics. 

On the other hand, our Hamiltonian formalism has permitted us to prove that Darboux transformations, both in the axial and polar sectors, are in one-to-one correspondence with the subclass of canonical transformations between canonical pairs describing master functions that have a perturbative Hamiltonian of the form of a harmonic oscillator subject to a background potential (with a suitable redefinition of the evolution parameter). This result had been demonstrated for axial gauge invariants in Ref. \cite{MGAC} and has been extended here to the polar case. We note that both the Darboux transformations and our canonical transformations form a group, in both cases with inverse. From our Hamiltonian perspective, the symmetry hidden in the Darboux transformations is just a canonical symmetry for a dynamical evolution which is timelike only inside the BH. In addition, we have also proved for the first time that, in our Hamiltonian description, there naturally appear transformations that mix polar and axial polarities if we do not impose their preservation. Such transformations include Chandrasekhar's transformation as a particular case \cite{BHM}.  
An issue that deserves further investigation is the analysis of the backreaction of the perturbations. It is worth remarking that our formalism contains backreaction in a natural manner up to quadratic contributions of the perturbations in the action. In fact, we have also seen that the zero modes describing the background receive quadratic corrections in the perturbations to maintain the canonical structure. Clearly, these corrections already provide background contributions. In addition, the zero modes have been treated exactly up to quadratic order in the perturbations when constructing the total Hamiltonian constraint, which restricts them together with the gauge invariant perturbations. Therefore, this constraint contains the additional missing backreaction that completes the correction of the background solutions when perturbations are included. 

Another interesting question is isospectrality. It is known that it holds for Schwarzschild solutions in general relativity, when one considers perturbations related by Darboux transformations. An appealing front for future research is to investigate whether our generalized Darboux transformations (in the sense that they are defined off shell) allow us to elucidate whether this isospectrality is as well valid for (some family of) effective backgrounds beyond Schwarzschild. A related, but in principle different question, is to study whether the canonical transformations corresponding to the Darboux ones can be implemented unitarily in any Fock representation of the perturbations, such as those proposed in the context of the hybrid quantization of the system \cite{MGA}. This unitarity goes beyond the issue of isospectrality, because it involves the whole infinite tower of possible values of $\ell$, the integer angular-momentum label. We expect that techniques similar to those discussed for perturbative fields on KS geometries in Refs. \cite{Alvaro1, Alvaro2} can allow us to resolve this question. 

\acknowledgments

G.A.M.M. and A.M.-S. are supported by MCIN/AEI/10.13019/501100011033 (Spanish Ministry of Science and Innovation) and FSE+ under the Grant No. PID2023-149018NB-C41. C.F.S. is supported by the Grant No. PID2022-137674NB-I00 from MCIN/AEI/10.13039/501100011033 (Spanish Ministry of Science and Innovation) and 2021-SGR-01529 (AGAUR, Generalitat de Catalunya). C.F.S. is also partially supported by the program Unidad de Excelencia Mar\'{\i}a de Maeztu CEX2020-001058-M (Spanish Ministry of Science and Innovation). A.M.-S. acknowledges support from the PIPF-2023 fellowship from Comunidad Aut\'onoma de Madrid. The reference number is PIPF-2023/TEC-30167.

G.A.M.M. dedicates this work to the memory of Prof. Jerzy Lewandowski, his friend and research colleague. {\sl {Even when time fails, evolution continues}}. 

\appendix

\section{Expressions for the perturbations \label{AppA}}

In this appendix, we provide the expressions of the perturbative gauge constraints and perturbative Hamiltonians presented in Sec. \ref{SecIII} and some formulas omitted in Sec. \ref{SecIV}. We use the same notation as in the main text of this
work.

For the axial sector, the perturbative constraint takes the form \cite{MM}
\begin{equation}
\label{eq: appA-A.1}
C_1[k_1^{\mathfrak{n}}] = \sum_{\mathfrak{n}} k_1^{\mathfrak{-n}}\left[ \frac{(\ell+2)!}{(\ell-2)!}(\Omega_b + \Omega_c)\frac{h_2^{\mathfrak{-n}}}{p_c^2} - 2p_2^{\mathfrak{-n}} + \frac{2\pi n}{L_0} \left( p_1^{\mathfrak{n}} - 4\ell(\ell+1)\frac{L_0^2}{p_b^2}\Omega_b h_1^{\mathfrak{n}} \right)\right],
\end{equation}
where the notation $\mathfrak{-n}$ denotes $\mathfrak{n}$ with a change in the sign of $n$, $k_1^{\mathfrak{n}}$ is the associated perturbative Lagrange multiplier, and $\omega_n=2\pi |n|/L_0$. Terms proportional to $\omega_n$ as well as sign reversals in some labels, arise from derivatives along the $\zeta$-direction. In addition, the axial perturbative Hamiltonian is \cite{MM}
\begin{equation}
\label{eq: appA-A.2}
\begin{aligned}
\,^{(ax)}\tilde{H}[\tilde{N}] &= \sum_{\mathfrak{n}}\frac{\tilde{N}}{2\kappa} \Bigg[\frac{p_b^2}{L_0^2}\frac{[p_1^{\mathfrak{n}}]^2}{\ell(\ell+1)} + \left(6\Omega_b^2 + 4\Omega_b\Omega_c + \frac{p_b^2}{L_0^2}\ell(\ell+1)\right)\frac{L_0^2}{p_b^2}\ell(\ell+1)[h_1^{\mathfrak{n}}]^2 - 4\Omega_b h_1^{\mathfrak{n}}p_1^{\mathfrak{n}}  - 4\Omega_c h_2^{\mathfrak{n}}p_2^{\mathfrak{n}}\\
&+ 4p_c^2\frac{(\ell-2)!}{(\ell+2)!}[p_2^{\mathfrak{n}}]^2 +\frac{(\ell+2)!}{4(\ell-2)!p_c^2}\left(2\Omega_b^2 + 4\Omega_c^2 + 4\Omega_b\Omega_c + \omega_n^2 p_c^2\right)[h_2^{\mathfrak{n}}]^2 + \frac{2\pi n (\ell+2)!}{L_0(\ell-2)!}h_1^{\mathfrak{n}}h_2^{\mathfrak{-n}}\Bigg].
\end{aligned}
\end{equation}

In turn, the perturbative gauge constraints for the polar sector are given by \cite{MM2,MGA}
\begin{equation}
\label{eq: appA-A.3}
\begin{aligned}
&\begin{aligned}
C_2[k_2^{\mathfrak{n}}] &= -\sum_{\mathfrak{n}} \tilde{N}k_2^{\mathfrak{n}}\Bigg[\Omega_b|p_c|p_3^{\mathfrak{n}} - \left(\omega_n^2 p_c^2 + 2\Omega_b^2 + 2\Omega_b\Omega_c + \frac{\ell^2+\ell+2}{2}\frac{p_b^2}{L_0^2}  + 4\frac{\tilde{H}_{\text{KS}}}{L_0}\right)\frac{h_3^{\mathfrak{n}}}{|p_c|} + \frac{p_b^2}{L_0^2|p_c|}\Omega_c p_6^{\mathfrak{n}}\\
&- \frac{L_0^2|p_c|}{p_b^2}\left(2\Omega_b\Omega_c + \frac{\ell^2+\ell+2}{2}\frac{p_b^2}{L_0^2} + 2\frac{\tilde{H}_{\text{KS}}}{L_0}\right)h_6^{\mathfrak{n}} + \frac{2\pi n}{L_0}|p_c|\ell(\ell+1)h_5^{\mathfrak{-n}} - \frac{(\ell+2)!}{4(\ell-2)!}\frac{p_b^2}{L_0^2|p_c|}h_4^{\mathfrak{n}}\Bigg],
\end{aligned}\\
&\begin{aligned}
C_3[k_3^{\mathfrak{n}}] &= \sum_{\mathfrak{n}} k_3^{\mathfrak{n}}\Bigg[\ell(\ell+1)\left(\frac{2\pi n}{L_0}|p_c|\left[4\frac{L_0^2}{p_b^2}\Omega_b h_5^{\mathfrak{-n}} - \frac{p_5^{\mathfrak{-n}}}{\ell(\ell+1)}\right]- 2\frac{L_0^2}{p_b^2}|p_c| \Omega_b  h_6^{\mathfrak{n}} - |p_c| p_3^{\mathfrak{n}}\right) + 2 |p_c| p_4^{\mathfrak{n}}\\
&- \frac{(\ell+2)!}{(\ell-2)!|p_c|}(\Omega_b+\Omega_c)h_4^{\mathfrak{n}}\Bigg],
\end{aligned}\\
&\begin{aligned}
C_4[k_4^{\mathfrak{n}}] &= \sum_{\mathfrak{n}} k_4^{\mathfrak{n}}\Bigg[\frac{4\pi n}{L_0}\left(\frac{L_0^2}{p_b^2}p_c^2\Omega_bh_6^{\mathfrak{-n}} - (\Omega_b+\Omega_c)h_3^{\mathfrak{-n}} - \frac{p_b^2}{L_0^2} p_6^{\mathfrak{-n}}\right) - 2\ell(\ell+1)(\Omega_b+\Omega_c)h_5^{\mathfrak{n}}+ \frac{p_b^2}{L_0^2}p_5^{\mathfrak{n}}\Bigg].
\end{aligned}
\end{aligned}
\end{equation}
Here, $k_I^{\mathfrak{n}}$ for $I=2,3,4$ are the corresponding Lagrange multipliers. Finally, the polar perturbative Hamiltonian is \cite{MM2,MGA}
\begin{equation}
\label{eq: appA-A.4}
\begin{aligned}
\,^{(po)}\tilde{H}[\tilde{N}] &= \sum_{\mathfrak{n}}\frac{\tilde{N}}{2\kappa}\Bigg[  \left(-\omega_n^2p_c^2+8\Omega_b(\Omega_b+\Omega_c)+4\frac{p_b^2}{L_0^2}+8\frac{\tilde{H}_{\text{KS}}}{L_0}\right)\frac{[h_3^{\mathfrak{n}}]^2}{p_c^2} - (4\Omega_b-2\Omega_c)h_6^{\mathfrak{n}}p_6^{\mathfrak{n}} + 2\frac{L_0^2}{p_b^2}p_c^2\Omega_bh_6^{\mathfrak{n}}p_3^{\mathfrak{n}}\\
&+ \frac{p_b^4}{L_0^4}\frac{[p_6^{\mathfrak{n}}]^2}{p_c^2} + \left(4\Omega_b(\Omega_b+\Omega_c)+2\frac{p_b^2}{L_0^2}+3\frac{\tilde{H}_{\text{KS}}}{L_0}\right)\frac{L_0^4}{p_b^4}p_c^2[h_6^{\mathfrak{n}}]^2 + 2\frac{p_b^2}{L_0^2}\left(\frac{2}{p_c^2}\Omega_b h_3^{\mathfrak{n}} - p_3^{\mathfrak{n}}\right)p_6^{\mathfrak{n}}\\
&- \frac{L_0^2}{p_b^2}\left(4\Omega_b(\Omega_b-\Omega_c)+(\ell+2)(\ell-1)\frac{p_b^2}{L_0^2}-4\frac{\tilde{H}_{\text{KS}}}{L_0}\right)h_3^{\mathfrak{n}}h_6^{\mathfrak{n}} +\frac{4\pi n}{L_0}\ell(\ell+1)h_3^{\mathfrak{n}}h_5^{\mathfrak{-n}} + \frac{p_b^2}{L_0^2}\frac{[p_5^{\mathfrak{n}}]^2}{\ell(\ell+1)}\\
&+ 2\frac{L_0^2}{p_b^2}\ell(\ell+1)\left(4\Omega_b(\Omega_b+\Omega_c)+\frac{p_b^2}{L_0^2}+2\frac{\tilde{H}_{\text{KS}}}{L_0}\right)[h_5^{\mathfrak{n}}]^2 - 4\Omega_b h_5^{\mathfrak{n}}p_5^{\mathfrak{n}} + 4p_c^2\frac{(\ell-2)!}{(\ell+2)!}[p_4^{\mathfrak{n}}]^2\\
&+ \frac{1}{4p_c^2}\frac{(\ell+2)!}{(\ell-2)!}\left(4(\Omega_b+\Omega_c)^2+2\frac{p_b^2}{L_0^2}+\omega_n^2p_c^2+4\frac{\tilde{H}_{\text{KS}}}{L_0}\right)[h_4^{\mathfrak{n}}]^2 - \frac{1}{2}\frac{(\ell+2)!}{(\ell-2)!}h_6^{\mathfrak{n}}h_4^{\mathfrak{n}} - 4\Omega_c h_4^{\mathfrak{n}}p_4^{\mathfrak{n}}\\
&- \frac{2\pi n}{L_0}\frac{(\ell+2)!}{(\ell-2)!}h_5^{\mathfrak{n}}h_4^{\mathfrak{-n}}\bigg].
\end{aligned}
\end{equation}

We now present the steps omitted in Sec. \ref{SecIV} that lead to the introduction of perturbative gauge invariant variables. For the axial sector, we perform the canonical transformation \cite{MM}
\begin{equation}
\label{eq: appB-B.1}
\begin{aligned}
&h_1^{\mathfrak{n}} = P_1^{\mathfrak{n}} - \frac{\pi n}{L_0}Q_2^{\mathfrak{-n}},\quad p_1^{\mathfrak{n}} = -Q_1^{\mathfrak{n}} - \ell(\ell+1)\frac{4\pi n L_0}{p_b^2}\Omega_b Q_2^{\mathfrak{-n}},\\
&h_2^{\mathfrak{n}} = Q_2^{\mathfrak{n}},\quad
p_2^{\mathfrak{n}} = P_2^{\mathfrak{n}} + \frac{\pi n}{L_0}Q_1^{\mathfrak{-n}} + \ell(\ell+1)\frac{4\pi n L_0}{p_b^2}\Omega_b P_1^{\mathfrak{-n}} + \frac{(\ell+2)!}{2(\ell-2)!p_c^2}(\Omega_b+\Omega_c)Q_2^{\mathfrak{n}}.
\end{aligned}
\end{equation}
This transformation makes $P_2^{\mathfrak{n}}$ equal to the axial perturbative constraint. 

The background dependence of the above transformation leads to a new Hamiltonian with additional contributions, which account for the change in the dynamics attributed to the perturbations under our new choice of variables. These contributions can be computed as $\chi$-derivatives of any generating function for the transformation. Recall that, for functions on the background phase space, this $\chi$-derivative is equal to the Poisson brackets with the background Hamiltonian $\tilde{H}_{KS}[\tilde{N}]$ (taken with respect to the background symplectic structure). This, combined with the modifications in the functional dependence owing to the change of perturbative variables itself, are the ingredients necessary to calculate the expression of the new perturbative Hamiltonian for the axial modes. Using the same notation as in Eq. \eqref{cuadraHami} for the coefficients of the different types of quadratic perturbative terms in this axial Hamiltonian, it is possible to show that \cite{MM2,MGAC}
\begin{equation}
\label{eq: appB-B.5}
\begin{aligned}
&\begin{aligned}
A_{(ax)} = \frac{1}{\ell(\ell+1)}\left[\frac{p_b^2}{L_0^2}+\frac{\omega_n^2p_c^2}{(\ell+2)(\ell-1)}\right], 
\end{aligned}\\
&\begin{aligned}
B_{(ax)} = \ell(\ell+1)\frac{L_0^2}{p_b^2}\left[8\Omega_b^2 + 8\Omega_b\Omega_c + (\ell^2+\ell+2)\frac{p_b^2}{L_0^2} + \frac{16\omega_n^2}{(\ell+2)(\ell-1)}\frac{L_0^2p_c^2}{p_b^2}\Omega_b^2\right],
\end{aligned}\\
&\begin{aligned}
C_{(ax)} = \left[4+\frac{8\omega_n^2}{(\ell+2)(\ell-1)}\frac{L_0^2p_c^2}{p_b^2}\right]\Omega_b .
\end{aligned}
\end{aligned}
\end{equation}

For the polar sector we similarly implement a background-dependent canonical transformation. The part corresponding to the configuration polar perturbative variables is
\begin{equation}
\label{eq: appB-B.6}
\begin{aligned}
&h_3^{\mathfrak{n}} = |p_c|\left[Q_3^{\mathfrak{n}} + \Omega_b Q_4^{\mathfrak{n}} -\frac{\ell(\ell+1)}{2} Q_5^{\mathfrak{n}}\right],\\
&h_4^{\mathfrak{n}} = |p_c| Q_5^{\mathfrak{n}},\\
&h_5^{\mathfrak{n}} = \frac{p_b^2}{L_0^2}Q_6^{\mathfrak{n}} + \frac{\pi n }{L_0}|p_c|Q_5^{\mathfrak{-n}},\\
&h_6^{\mathfrak{n}} = \frac{p_b^2}{L_0^2|p_c|}\left[\Omega_c Q_4^{\mathfrak{n}} +  \frac{L_0^2}{p_b^2}\frac{2}{\Lambda_{\ell}}(\Omega_b P_3^{\mathfrak{n}} - P_4^{\mathfrak{n}}) + \frac{4\pi n}{L_0}|p_c|Q_6^{\mathfrak{-n}}\right].
\end{aligned}
\end{equation}
Recall that $\Lambda_{\ell} = (\ell+2)(\ell-1) - 6L_0^2\Omega_b\Omega_c /p_b^2$. For the momenta, the transformation of the perturbative mode variables is \cite{MM2,MGA}
\begin{equation}
\label{eq: appB-B.7}
\begin{aligned}
&\begin{aligned}
p_3^{\mathfrak{n}} &= \frac{1}{|p_c|}\Bigg[\left(1-\frac{4L_0^2}{p_b^2}\frac{\Omega_b^2}{\Lambda_{\ell}}\right)P_3^{\mathfrak{n}} + \left((\ell+2)(\ell-1)\frac{p_b^2}{2L_0^2} + \omega_n^2p_c^2\right)\frac{1}{\Omega_b} Q_3^{\mathfrak{n}} + \left(\frac{p_b^2}{2L_0^2} \Lambda_{\ell}+ \omega_n^2p_c^2 + \Omega_b\Omega_c\right)Q_4^{\mathfrak{n}}\\
&+\frac{4L_0^2}{p_b^2}\frac{\Omega_b}{\Lambda_{\ell}}P_4^{\mathfrak{n}} - \frac{4\pi n}{L_0}|p_c|(\Omega_b+\Omega_c)Q_6^{\mathfrak{-n}}\bigg],
\end{aligned}\\
&\begin{aligned}
p_4^{\mathfrak{n}} &= \frac{1}{|p_c|}P_5^{\mathfrak{n}} - \frac{4\pi n}{L_0}(\Omega_b+\Omega_c)Q_6^{\mathfrak{-n}} + \left[\frac{\ell(\ell+1)}{2|p_c|}\left((\ell+2)(\ell-1)\frac{p_b^2}{2L_0^2} + \omega_n^2p_c^2\right)\frac{1}{\Omega_b} - (\Omega_b-\Omega_c)\right]Q_3^{\mathfrak{n}}\\
&+ \frac{\pi n L_0}{p_b^2} P_6^{\mathfrak{-n}} + \frac{L_0^2}{p_b^2 |p_c|}\left[\omega_n^2 p_c^2 \left(\Omega_b^2-\Omega_b\Omega_c-\frac{p_b^2}{2L_0^2}\Lambda_{\ell}\right)+\frac{\ell(\ell+1)}{4}\frac{p_b^2}{L_0^2}\left((\ell+2)(\ell-1)\frac{p_b^2}{L_0^2} + 2 \omega_n^2p_c^2\right)\right]Q_4^{\mathfrak{n}}\\
&-\frac{L_0^2}{p_b^2} \left[1+\omega_n^2\frac{L_0^2}{p_b^2}|p_c|\right]\frac{2\Omega_b }{\Lambda_{\ell}} P_4^{\mathfrak{n}} + \left[\frac{\ell(\ell+1)}{2|p_c|} + \frac{L_0^2}{p_b^2}\left(1+\omega_n^2\frac{L_0^2}{p_b^2}|p_c|\right)\frac{2\Omega_b^2}{\Lambda_{\ell}}\right]P_3^{\mathfrak{n}} + \frac{(\ell+2)!}{(\ell-2)!}\frac{\Omega_b+\Omega_c}{2|p_c|}Q_5^{\mathfrak{n}},
\end{aligned}\\
&\begin{aligned}
p_5^{\mathfrak{n}} &= \frac{L_0^2}{p_b^2}P_6^{\mathfrak{n}} - \frac{4\pi n L_0}{p_b^2}|p_c|\left[(\Omega_b-\Omega_c)Q_3^{\mathfrak{-n}} - \left(\frac{L_0^2}{p_b^2} \frac{2}{\Lambda_{\ell}}- \Omega_b^2 + \Omega_b\Omega_c\right)Q_4^{\mathfrak{-n}} - \ell(\ell+1)\Omega_b Q_5^{\mathfrak{-n}}\right]\\
&+ 2(\ell+2)(\ell-1)(\Omega_b+\Omega_c) Q_6^{\mathfrak{n}},
\end{aligned}\\
&\begin{aligned}
p_6^{\mathfrak{n}} &= \frac{L_0^2}{p_b^2}|p_c|\left[\frac{L_0^2}{p_b^2}\frac{2\Omega_b}{\Lambda_{\ell}}(\Omega_b P_3^{\mathfrak{n}} - P_4^{\mathfrak{n}}) +\!\left(\frac{p_b^2}{L_0^2} \frac{\Lambda_{\ell}}{2}+ \Omega_b\Omega_c - 2\Omega_b^2\right)\! Q_4^{\mathfrak{n}} - 2\Omega_b Q_3^{\mathfrak{n}} + \ell(\ell+1)\Omega_b Q_5^{\mathfrak{n}}+ \frac{4\pi n}{L_0} |p_c|\Omega_b Q_6^{\mathfrak{-n}}\right].
\end{aligned}
\end{aligned}
\end{equation}
With this transformation, the momenta (modes) $P_I^{\mathfrak{n}}$ for $I=4$, $5$, and $6$, become equal to the polar Abelianized gauge constraints, which differ from those obtained in the polar sector by linearization of the spatial diffeomorphism constraints of general relativity only by a redefinition of $C_2[k_2]$, namely
\begin{equation}
\label{Abelianredef}
\tilde{C}_2[k_2^{\mathfrak{n}}] = C_2[k_2^{\mathfrak{n}}] + \sum_{\mathfrak{n}}2\tilde{N}k_2^{\mathfrak{n}}\left[2Q_3^{\mathfrak{n}} + (2\Omega_b+\Omega_c)Q_4^{\mathfrak{n}} + \frac{L_0^2}{p_b^2}\frac{2}{\Lambda_{\ell}}(\Omega_b P_3^{\mathfrak{n}} - P_4^{\mathfrak{n}})\right]\frac{\tilde{H}_{\text{KS}}}{L_0}.
\end{equation}
At our quadratic perturbative truncation of the action, this is also a constraint, linear in the perturbations, that can be used in place of $C_2[k_2]$ and such that it commutes with the rest of polar gauge constraints under the Poisson structure for the perturbations (with the background kept fixed). This substitution, as well as the change to our new perturbative variables, must be accompanied by a redefinition of Lagrange multipliers and of the lapse function $\tilde{N}$ that has been explained in the main text and for which we do not give explicit expressions here. They can be consulted in Refs. \cite{MM,MGAC}. 

Finally, by the same reasons explained when we discussed the canonical transformation of the axial mode variables, the polar contribution to the perturbative Hamiltonian changes in the above background-dependent canonical transformation. The coefficients of the quadratic perturbative terms in the polar sector are, with the notation of Eq. \eqref{cuadraHami}, given by \cite{MM2,MGA}
\begin{equation}
\label{eq: appB-B.14}
\begin{aligned}
&\begin{aligned}
A_{(po)} &= \frac{4\omega_n^2 p_c^2}{\ell(\ell+1)}\frac{L_0^4}{p_b^4}\left(\frac{p_b^2}{L_0^2} + \frac{\omega_n^2 p_c^2}{(\ell+2)(\ell-1)}\right) (\Omega_b-\Omega_c)^2+ \frac{(\ell+2)!}{(\ell-2)!}\left(\frac{p_b^2}{2L_0^2}+ \frac{\omega_n^2 p_c^2}{(\ell+2)(\ell-1)}\right)^2 \frac{1}{\Omega_b^2}\\
&+ \left(\frac{p_b^2}{2L_0^2}+ \frac{\omega_n^2 p_c^2}{(\ell+2)(\ell-1)}\right)\frac{4\omega_n^2 L_0^2  p_c^2}{p_b^2\,\Omega_b}(\Omega_b-\Omega_c)- \omega_n^2 p_c^2 + \left(\frac{(\ell+2)(\ell-1)}{2}\frac{p_b^2}{L_0^2}+ \omega_n^2 p_c^2\right)\frac{p_b^2}{\Omega_b^2L_0^2}\\
&+ \left((\ell+2)(\ell-1)\frac{p_b^2}{L_0^2} (\Omega_b + \Omega_c) + 4\omega_n^2 p_c^2\Omega_b\right)\frac{1}{\Omega_b},
\end{aligned}\\
&\begin{aligned}
B_{(po)} &= \frac{\ell(\ell+1)}{(\ell+2)(\ell-1)},
\end{aligned}\\
&\begin{aligned}
C_{(po)} &= \ell(\ell+1) \left(\frac{p_b^2}{2L_0^2}+ \frac{\omega_n^2 p_c^2}{(\ell+2)(\ell-1)}\right)\frac{2}{\Omega_b}+ \frac{4\omega_n^2 }{(\ell+2)(\ell-1)}\frac{L_0^2p_c^2 }{p_b^2}(\Omega_b-\Omega_c) - 2(\ell+2)(\ell-1)\frac{\Omega_b}{\Lambda_{\ell}}.
        \end{aligned}
    \end{aligned}
\end{equation}

With the axial and polar canonical transformations that we have introduced in this appendix, the gauge invariant modes are isolated and described by the canonical pairs of mode variables $(Q_1^{\mathfrak{n}},P_1^{\mathfrak{n}})$ and $(Q_3^{\mathfrak{n}},P_3^{\mathfrak{n}})$, respectively for the axial and polar degrees of freedom. They are subject only to one constraint: the total Hamiltonian constraint obtained by adding the axial and polar perturbative contributions that we have specified above and the background contribution provided by the background Hamiltonian $\tilde{H}_{KS}[\tilde{N}]$, though with the background phase-space variables evaluated at their corrected values, in which quadratic perturbative modifications restore the global canonical structure of the combined system formed by them and the gauge invariant perturbations. We analyze such modifications in the next appendix.

\section{Perturbative corrections to the background variables \label{AppB}}

In this appendix we explain how to modify the background phase-space variables with quadratic perturbative corrections so that, at out order of perturbative truncation of the action, the description of the combined system formed by the background and the perturbations remains canonical. 

Our analysis is valid for any background-dependent canonical transformation of the perturbations that initially treats the background as fixed. A transformation of this type can always be completed into a canonical transformation for both the perturbative and background sectors, if perturbations are truncated at second order in the action, therefore neglecting cubic and higher-order contributions. Since our discussion is general, it is convenient to introduce an abstract compact notation. Thus, let us call the original background variables and perturbative mode variables respectively as $\{x^j\} = \{x^j_q, x^j_p\}$ and $\{X_J^{\mathfrak{n}}\} = \{X_{q_{J}}^{\mathfrak{n}}, X_{p_{J}}^{\mathfrak{n}}\}$, where the indices $j$ and $J$ respectively label background and perturbative degrees of freedom, and the subscripts $q$ and $p$ denote configuration and momentum variables. These original variables form a canonical set. We then consider a canonical transformation on the perturbative sector, keeping the background nondynamical, to new perturbative variables $\{W_J^{\mathfrak{n}}\} $, given by the inverse of the relations $X_J^{\mathfrak{n}}=X_J^{\mathfrak{n}} [x^j,W_J^{\mathfrak{n}}]$. A new canonical set, which restores the dynamical character of the background, is obtained at our truncation order through the change to new background variables $\{w^j\} = \{w^j_q, w^j_p\}$ given by \cite{LMM}
\begin{equation}
\label{backcorrections}
\begin{aligned}
&w^j_{q} = x^j_{q} + \frac{1}{2}\sum_{\mathfrak{n}}\sum_{J} \left[ X^{\mathfrak{n}}_{q_J} \frac{\partial X^{\mathfrak{n}}_{p_J}}{\partial x^j_{p}} - \frac{\partial X^{\mathfrak{n}}_{q_J}}{\partial x^j_{p}}X^{\mathfrak{n}}_{p_J}\right], \\ 
&w^j_{p} = x^j_{p} - \frac{1}{2}\sum_{\mathfrak{n}}\sum_{J} \left[ X^{\mathfrak{n}}_{q_J} \frac{\partial X^{\mathfrak{n}}_{p_J}}{\partial x^j_{q}} - \frac{\partial X^{\mathfrak{n}}_{q_J}}{\partial x^j_{q}}X^{\mathfrak{n}}_{p_J}\right].
\end{aligned}
\end{equation}

\end{document}